\documentclass[sort&compress]{elsarticle}

\usepackage{lmodern}

\usepackage[left=1.20in,right=1.20in,top=1.0in,bottom=1.0in]{geometry}
\usepackage{amsmath,amssymb,mathtools,bm,isomath,amsthm,cancel}
\usepackage{enumitem}

\usepackage{color,graphicx,epstopdf}  
\graphicspath{./Figures/}

\usepackage{tikz,pgfplots}
\usetikzlibrary{plotmarks,calc,arrows.meta,arrows,intersections,quotes,angles,backgrounds,patterns,shapes,fit,positioning}
\usepgfplotslibrary{patchplots,colormaps,statistics,fillbetween}
\pgfplotsset{compat=1.16}
\usepackage{grffile}
\pgfplotsset{every axis/.append style={
        scaled ticks = false, 
        tick label style={/pgf/number format/fixed}}}
        
\usetikzlibrary{external}
\newlength\figureheight 
\newlength\figurewidth 
\newlength\imagewidth

\usepackage{parskip}
\usepackage{booktabs,multirow}
\usepackage[linesnumbered,ruled,vlined]{algorithm2e}

\newcommand{\real}{\mathbb{R}} 
\newcommand{\diff}{\mathrm{d}}

\newcommand{\density}{f}
\newcommand{\vel}{v} 
 
\newcommand{\conc}{\phi} 
\DeclareMathOperator{\Prob}{Pr}
\DeclareMathOperator{\Exp}{\mathbb{E}}

\DeclareMathOperator{\Cov}{Cov}

\DeclareMathOperator{\Tr}{Tr}

\DeclareMathOperator*{\argmax}{argmax}
\DeclareMathOperator*{\argmin}{argmin}
\DeclarePairedDelimiter\dual{\langle}{\rangle}
\DeclarePairedDelimiter\abs{\lvert}{\rvert}

\DeclarePairedDelimiter\norm{\lVert}{\rVert}

\definecolor{darkblue}{RGB}{0,0,190}
\definecolor{darkred}{RGB}{190,0,0}
\definecolor{darkgreen}{RGB}{0,120,0}

\pgfplotscreateplotcyclelist{eigenvectors}{blue,densely dashed,mark=star\\
red,densely dashed,mark=pentagon*\\
black,densely dashed,mark=diamond*\\
blue,mark=triangle*\\
red,mark=square*\\
black,mark=*\\
}

\begin{document}

\title{Optimal Design of Validation Experiments for the Prediction of Quantities of Interest}


\author[1]{Antonin Paquette-Rufiange\corref{cor1}}
\ead{antonin.paquette-rufiange@polymtl.ca}
\author[1]{Serge Prudhomme}
\author[1]{Marc Laforest}
\address[1]{Département de mathématiques et de génie industriel,  Polytechnique Montréal,
H3T 1J4, Montréal, Canada}

\cortext[cor1]{Corresponding author}

\begin{abstract} 
Numerical predictions of quantities of interest measured within physical systems rely on the use of mathematical models that should be validated, or at best, not invalidated. Model validation usually involves the comparison of experimental data (outputs from the system of interest) and model predictions, both obtained at a specific validation scenario. The design of this validation experiment should be directly relevant to the objective of the model, that of predicting a quantity of interest at a prediction scenario. In this paper, we address two specific issues arising when designing validation experiments. The first issue consists in determining an appropriate validation scenario in cases where the prediction scenario cannot be carried out in a controlled environment. The second issue concerns the selection of observations when the quantity of interest cannot be readily observed. The proposed methodology involves the computation of influence matrices that characterize the response surface of given model functionals. Minimization of the distance between influence matrices allow one for selecting a validation experiment most representative of the prediction scenario. We illustrate our approach on two numerical examples. The first example considers the validation of a simple model based on an ordinary differential equation governing an object in free fall to put in evidence the importance of the choice of the validation experiment. The second numerical experiment focuses on the transport of a pollutant and demonstrates the impact that the choice of the quantity of interest has on the validation experiment to be performed.     
\end{abstract}



\maketitle

\section{Introduction}

The development of models to describe a system of interest (be it physical, biological, or societal) is central in understanding and predicting its behavior. Mathematical models may stem from physical laws and/or empirical considerations for either some parts of the system or the whole system itself. In order to assess the potential of a model to explain and predict a given quantity of interest (QoI), it must be validated, or at best, not invalidated. In other words, the error between the model and the reality it is supposed to describe must be precisely quantified with respect to the QoI. The validation process implies comparing outputs of the model with experimental data acquired by testing the system of interest. Over the past decades, several procedures for model validation have been proposed in the literature. For instance, Oberkampf and Trucano~\cite{oberkampf_verification_2008} present a detailed overview of verification and validation processes while Roy and Oberkampf~\cite{roy_comprehensive_2011} provide a thorough discussion on these two processes with a focus on the treatment of both aleatory and epistemic uncertainty. A recent review article by Riedmaier et al.~\cite{riedmaier_unified_2021} discusses in details the scope and shortfalls of existing approaches for validation and prediction. The paper of Lee et al.~\cite{lee_review_2019} provides an extensive review of the literature addressing the problems of uncertainty propagation, parameters calibration, and model validation. Model validation has become increasingly relevant and primordial in the field of computational sciences and engineering due to the complexity of scientific models currently employed and the demand for predictive and quantitative results. However, a critical aspect remains the selection of suitable validation experiments, since they provide the experimental data against which the model prediction is compared and thus directly influence the decision whether the model is deemed valid or not. This issue is even more important when the number of validation experiments or amount of data are limited. Yet, the precise design of such validation experiments is still a question largely overlooked in the literature, with the exception of a few contributions. 

Oliver et al.~\cite{oliver_validating_2015} stress the importance of the validation process for predictive modeling, even more so in the case of models made of several coupled sub-models. They essentially identify two fundamental issues when tackling model validation: 1) that of validating a model if one cannot reproduce the conditions under which the predictions are made; 2) that of validating a model if the QoI one wishes to predict cannot be observed in practice. To answer these two questions, the authors estimate the modeling errors affecting the sub-models and propagate them to the QoI at the prediction scenario. They then propose a series of guidelines to ensure that the calibration and validation experiments are relevant for prediction purposes~\cite[Section 3.3]{oliver_validating_2015}. These guidelines can be roughly summarized as follows: if the QoI is sensitive to certain model parameters and/or certain modeling errors, then the calibration and validation experiments should reflect these sensitivities. However, the discussion remains essentially qualitative and the analysis is performed a posteriori, that is, once calibration and validation experiments have been performed, they subsequently verify that these are actually relevant.

The paper of Hamilton and Hills~\cite{hamilton_relation_2010} and the subsequent report of Hills~\cite{hills_roll-up_2013} also address these aforementioned issues. The authors develop a meta-model allowing the extrapolation of the modeling error to the QoI at prediction scenario. More precisely, they identify a linear relationship between the sensitivities (some kind of elementary effects) of the QoI at prediction scenario to the sensitivities of the observables at validation scenario. Weights could then be attributed to more relevant validation observations with respect to the objective of the prediction. However, the assessment is again made a posteriori. Moreover, the authors do not address the design of validation experiments. We nevertheless note that Hills~\cite[Section 1.3]{hills_roll-up_2013} presents several cases where a validation experiment may fail to provide a useful assessment on whether the model can be employed to predict a QoI at a prediction scenario.

Other works have also studied the problem of validating mathematical models made of multiple sub-models. Among those, we mention the work of Hills and Leslie~\cite{hills_statistical_2003}, who employ local derivative-based indices to weight the importance of the validation experiments for the sub-models with respect to the numerical prediction using the full model. In the same vein, the work of Li and Mahadevan~\cite{li_role_2016} proposes an integrated approach to take into account the validity of the lower-level models and their relevance for the system-level model. More precisely, they employ a model reliability metric to assess the validity of lower-level models and perform a sensitivity analysis based on Sobol indices~\cite{sobol_global_2001} to quantify the relation of these same lower-level models to the model for the full system.

Farrell et al.~\cite{farrell_bayesian_2015} propose the so-called Occam-Plausibility algorithm that allows one for identifying the simplest and most plausible model among a given family of competing models that they rank in terms of the posterior model plausibility. The paper of Tan et al.~\cite{tan_toward_2022} enriches the Occam-Plausibility algorithm by providing two criteria about the selection of the validation experiment to be performed. The first criterion stipulates that the sensitivity indices of the validation scenario must be close to the sensitivity indices of the prediction scenario. They employ Sobol indices~\cite{sobol_global_2001} as their sensitivity indices, thus considering the decomposition of the variance as a notion of sensitivity. This criterion is also verified a posteriori, once a validation scenario has been proposed. The second criterion checks if the validation scenario provides an information gain on the posterior distribution of the model parameters compared to the calibration scenarios.

Ao et al.~\cite{ao_design_2017} specifically study the optimal design of validation experiments in the context of life prediction models. Their approach is actually similar to the Bayesian optimal design approach described in Section~\ref{ssec:calibration}. However, their methodology relies on a fine model representing the system of interest, thus necessitating additional a priori knowledge on the system of interest, which is not always available.

We also mention the paper of Sunseri et al.~\cite{sunseri_hyper-differential_2020} that investigates the influence of auxiliary parameters on the solution of inverse problems. Although their study does not address the problem of validation, their methodology is relevant to the present work. They determine some pointwise and global sensitivity indices using the gradient of the solution to the inverse problem with respect to the auxiliary variables. Whenever an auxiliary parameter possesses a high sensitivity index, it indicates that a small perturbation in its value significantly impacts the solution to the inverse problem and that some effort should be spent in order to decrease the uncertainty of this auxiliary parameter.  

The present work aims at refining some of the concepts laid out in the aforementioned papers, in particular, the use of sensitivity indices to assess the relevance of validation experiments. The main objective here is therefore to describe a novel methodology for the optimal design of validation experiments. The proposed approach essentially relies on the formulation of two distinct optimization problems, which are formulated in such a way that the behavior of the model under the validation conditions resembles as much as possible that of the same model, but under the prediction conditions. In this manner, one can gain confidence in the model predictions, in the sense that the choice of the validation experiment is tailored so as  to produce a validation scenario in which the observables express the same behavior as that of the QoI in the prediction scenario. In other words, the modeling errors observed in the validation scenario will be similar to the modeling errors that one would expect in the prediction scenario.

The paper is organized as follows.  Section~\ref{sec:background} introduces all the necessary concepts and notations to precisely describe the validation process. A detailed account of the different types of parameters and sources of uncertainty involved in a model will be provided. We also describe methodologies for calibration and validation and give a brief overview of methods for optimal design of calibration experiments. Section~\ref{sec:optimal_design_validation} presents our novel approach for the design of validation experiments tailored toward the prediction of a quantity of interest that cannot potentially be observed. We briefly review the Active Subspace method as the chosen method to perform the sensitivity analysis. Two optimal design problems will be formulated in order to compute the appropriate control and sensor parameters for the validation scenario. We then present some numerical examples in Section~\ref{sec:examples}. We first consider a toy problem, that of an object in free fall, to illustrate the validation process, with an emphasis on the importance of designing optimal validation experiments. We then study a second example, consisting of a simplified pollutant transport problem, to highlight the importance of the definition of the QoI in the design of optimal validation experiments. Section~\ref{sec:conclusion} provides some concluding remarks along with a series of open questions regarding the proposed methodology and how it relates to the larger process of uncertainty quantification and model validation.

\section{Terminology and Preliminary Notions about Validation and Predictive Modeling}
\label{sec:background}

The purpose of this section is to introduce some preliminary notions and notations in order to precisely describe the validation process. We recall that the primary objective of predictive modeling is to be able to obtain some quantitative  predictions regarding a system of interest. At first sight, the task may seem rather straightforward to achieve as it simply aims at evaluating the output of a model. However, as mentioned earlier, for numerical predictions to be of any usefulness, one should be confident that the model remains valid for its intended use and that the predictions obtained with the model are in fact accurate descriptions of the reality. It should be clear by now that the validity of the model should be assessed with respect to the quantities of interest that one would like  to predict using the model at specific conditions/regimes. In the presentation below, we will strive to use the same terminology as that commonly introduced in the literature, with a few exceptions in order to emphasize the different roles of some of the model parameters and model outputs.

\subsection{Abstract Model}
\label{ssec:abstract_model}

We define a generic \emph{deterministic model} as the triplet $(r,p,u)$, where $p=(p_1,\ldots,p_d)\in \mathcal{P}$ are the \emph{parameters},~$u = (u_1,\ldots,u_s)\in\mathcal{U}$ are the \emph{state variables}, and $r$ include the \emph{equilibrium relation} and initial and/or boundary conditions
\begin{align}
\label{eq:equilibrium}
r(p,u) = 0. 
\end{align}
For the sake of simplicity, the parameter space $\mathcal{P} \subseteq \real^d$ is assumed to be finite dimensional and the state space~$\mathcal{U}$ to be either a Banach or Hilbert space. The equilibrium relation~$r$ may involve ordinary or partial differential equations, algebraic equations, or any combination thereof. The equilibrium relation encapsulates the set of hypotheses and scientific understanding that hopefully  lead to a sufficiently accurate model of the system of interest. An important property that the equilibrium relation~$r$ must possess is well-posedness: for each $p\in \mathcal{P}$, there exists a unique and stable solution~$u\in\mathcal{U}$ satisfying~\eqref{eq:equilibrium}. In other words, there exists a function~$g$ such that $u=g(p)$. 

As an example, we consider a simple model that will be used in Section~\ref{ssec:example1} to illustrate our methodology. The model describes the motion of a spherical projectile launched vertically in ambient air at sea level. The evolution of the physical system can be described by the linear ordinary differential equation and initial conditions
\begin{subequations}
\label{eq:model_projectile}
\begin{align}
    m\frac{\diff^2 u}{\diff t^2}(t) + 3\pi\ell \exp(\mu) \frac{\diff u}{\diff t}(t)= -mg, &\qquad \forall t \in (0,T)\\
    u(0) = u_0, &\\
    \frac{\diff u}{\diff t}(0)  = v_0, &
\end{align}   
\end{subequations}
where~$u(t)$ is the altitude reached by the projectile at time~$t$, $m$ and~$\ell$ are its mass and diameter, $u_0$ and $v_0$ are its initial position and velocity, respectively, $g$ is the gravitational constant, and~$\mu$ is a viscosity parameter for a spherical object travelling in air~\cite[p.488]{white_fluid_2009}. In this example, the parameters of the problem are given by $p=(m,\ell,u_0,v_0,g,\exp(\mu))\in \mathcal{P} = (\real^+)^6$ and the state variable consists in the altitude $u \in \mathcal{U} = C^1(\real)$ equipped with the uniform norm\ on~$u$ and~$\diff u/\diff t$. The set of equilibrium relation and initial conditions can easily be obtained from the initial-value problem~\eqref{eq:model_projectile}. The model is constructed by making the following hypotheses regarding the physical system:
\begin{enumerate}[itemsep=0pt,topsep=2pt,parsep=0pt,leftmargin=25pt]
    \item The trajectory of the projectile is supposed to be one dimensional, so only forces acting vertically are taken into account.
    \item The friction force is assumed to be proportional to the velocity of the projectile.
    \item The viscosity of the air and the gravitational constant remain constant with respect to the altitude.
\end{enumerate}

\subsection{Classes of Parameters}

We split the parameters~$p$ of a problem into two distinct subsets, namely the \emph{control parameters}~$x$ and the \emph{model parameters}~$\theta$. We also consider an additional set of parameters that we call the \emph{sensor parameters}~$z \in \Omega$, where~$\Omega$ is the domain of definition of the state variable~$u$. Each type of parameters will play some specific roles, as described below. Similar definitions can be found in the papers cited in the introduction, even if no consensus has yet been reached in the literature on a common terminology.

The \emph{control parameters}~$x \in \mathcal{X}$ represent the parameters that one can control when performing an experiment on the system of interest. These parameters can be viewed as the set of inputs which uniquely determine the experimental scenario. Examples of such parameters are the parameters appearing in the initial conditions and boundary conditions, the geometrical data, and the physical properties directly measurable by an experimental apparatus (mass, length, temperature, etc.). For the example of Section~\ref{ssec:abstract_model}, the control parameters are given by the mass and diameter of the projectile as well as its initial altitude and velocity, hence~$x=(m,\ell,u_0,v_0)$.

The \emph{model parameters}~$\theta \in \mathcal T$ are the remaining parameters necessary for the full description of the model ($\theta = p\setminus x$). They are coined this way due to the fact that they explicitly stem from the assumptions and simplifications that one makes in defining a model to describe a particular phenomenon. Other hypotheses could potentially lead to a new model with different model parameters, but will not change the control parameters. Naturally, the definition of the parameters~$p$, and hence of the model parameters~$\theta$, depends directly on what a parameter would be deemed as such by the modeler and/or user. For example, numerical constants like~$\pi$ in~\eqref{eq:model_projectile} clearly do not constitute parameters, but the picture could be far less clear for other quantities. As a rule of thumb, if the value of a quantity may vary or if it is of any interest in the analysis of the model, then it can be considered as a model parameter. Examples of model parameters can be those that characterize the properties of a material (e.g.\ Young modulus, Poisson coefficient) or some dimensionless quantities (e.g.\ Reynolds number, Prandlt number). Model parameters are specified either by prior knowledge or by a calibration process. A more detailed discussion about the latter point will be presented in Section~\ref{ssec:uncertainty_errors}. In the case of the spherical projectile, the model parameters are~$\theta = (g,\mu)$ since they result from assumptions in the modeling process ($g$ is independent of the altitude and~$\mu$ is related to a sub-model for air friction). 

A final type of parameters are the so-called \emph{sensor parameters}~$z\in\Omega$. They often represent location defined in terms of the independent variables (e.g.\ space or time) upon which the state variable~$u$ would be observed. In the example of the projectile, the sensor parameters will be the time at which we wish to evaluate the altitude, so~$z=t$. 

\subsection{Classes of Scenarios}

The control and model parameters allow one to define the regime under which both the system and the model operate. A specific regime will be referred to as a \emph{scenario}. For the sake of simplicity here, we shall restrict the model parameters~$\theta$ to be the same for the validation and prediction scenarios. For the projectile example, this implies that the prediction and experimental apparatus be at sea level and in the ambient atmosphere (same gravitational constant~$g$ and same viscosity~$\mu$). In other words, only the control parameters~$x$ are needed to fully characterize a scenario. We shall study three classes of scenarios, with a particular emphasis on the first two.
\begin{enumerate}[itemsep=3pt,topsep=2pt,parsep=0pt,leftmargin=25pt]
    \item \textbf{Prediction Scenario:} Defines the conditions, described by~$x_{\text{pred}}$, under which predictions on the system of interest will be made.
    \item \textbf{Validation Scenario:} Defines the conditions, described by~$x_{\text{val}}$, under which the validation experiment will be carried out.
    \item \textbf{Calibration Scenario:} Defines the conditions, described by~$x_{\text{cal}}$, under which the calibration experiment will be performed. We note here that model calibration essentially aims at estimating the uncertainty of the model parameters~$\theta$.
\end{enumerate}

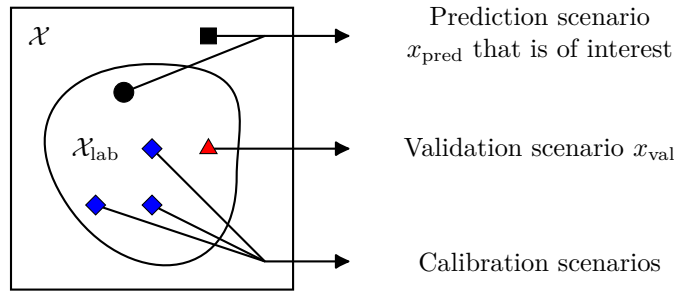
\begin{figure}[tbh]
    \centering
    \def\figurescale{0.75} 
    \tikzset{external/export next=false}
    \begin{tikzpicture}[scale=\figurescale]

\pgfmathsetlengthmacro{\twidth}{4cm*\figurescale}

\tikzset{dot/.style = {circle, minimum size=\figurescale*10pt, draw=black, fill=black, inner sep=0pt, outer sep=0pt},
         diamond_custom/.style = {diamond, minimum size=\figurescale*10pt, draw=black, fill=blue, inner sep=0pt, outer sep=0pt},
         triangle_custom/.style = {regular polygon, regular polygon sides=3, minimum size=\figurescale*10pt,
         draw=black, fill=red, inner sep=0pt, outer sep=0pt}
         }

\tikzstyle{arrow} = [thick,-{Latex[length=\figurescale*3mm,width=\figurescale*3mm,round]}]
\tikzstyle{custom text} = [text width=1.6*\twidth,text centered]

\draw [thick] (-1.5,0) .. controls (0.5,0) and (0,-1) .. (0,-2) .. controls (0,-4) and (-2,-4) .. (-3,-2.5) .. controls (-4,-1) and (-3,0) .. (-1.5,0) -- cycle ;

\draw [thick] (-4,1) rectangle (1,-4);

\node at (-3.5,0.5) {$\mathcal{X}$};
\node at (-2.5,-1.5) {$\mathcal{X}_{\text{lab}}$};

\node [fill=black,draw=black,rectangle, minimum size=\figurescale*8pt, inner sep=0pt, outer sep=0pt] (v7) at (-0.5,0.5) {};
\node [dot] (v6) at (-2,-0.5) {};

\node [triangle_custom] (v5) at (-0.5,-1.5) {};

\node [diamond_custom] (v1) at (-1.5,-1.5) {};
\node [diamond_custom] (v4) at (-1.5,-2.5) {};
\node [diamond_custom] (v2) at (-2.5,-2.5) {};

\coordinate (end arrow) at (2,0);

\draw [arrow] (v1) -- (0.5,-3.5) node [coordinate] (v3) {} -- (v3 -| end arrow) node[custom text,right]{Calibration scenarios};
\draw [thick] (v2) -- (v3);
\draw  [thick] (v4) edge (v3);
\draw [arrow] (v5) -- (v5 -| end arrow) node[custom text,right]{Validation scenario~$x_\text{val}$};
\draw [arrow] (v6) -- (0.5,0.5) node [coordinate] (v8) {} -- (v8 -| end arrow) node [custom text,right] {Prediction scenario $x_\text{pred}$ that is of interest};
\draw  [thick] (v7) -- (v8);
\end{tikzpicture} 
    \caption{Illustration of the various scenarios. The prediction scenario may be producible (black circle) or may not be producible experimentally (black square). The validation and calibration scenarios are elements of the set~$\mathcal{X}_\text{lab}$ of controlled environments.}
    \label{fig:parameter_set}
\end{figure}

On the one hand, the validation and calibration experiments are performed within a \emph{set} $\mathcal{X}_\text{lab} \subset \mathcal{X}$ of \emph{controlled environments}. On the other hand, the prediction scenario $x_\text{pred}$ may or may not be producible experimentally, so that $x_\text{pred}$ may not belong to $\mathcal{X}_\text{lab}$. Figure~\ref{fig:parameter_set} provides a conceptual illustration of the various scenarios. 

\subsection{Observables and Quantities of Interest}

We reiterate that a mathematical model constitutes only an abstraction of a physical system of interest. In fact, one has access to limited information about the system of interest. This information will be collectively referred to as the \emph{experimental observables}~$y_{\text{exp}}$. The experimental observables~$y_{\text{exp}}$ can be acquired through a variety of experimental devices and/or statistical surveys. Importantly, they are obtained for a particular scenario and for given sensor parameters, so that they are functions of $x$ and $z$, i.e.\ $y_{\text{exp}} = y_{\text{exp}} (x,z)$. We also assume that the model~$(r,p,u)$ is sufficiently rich so that each experimental observable can be abstracted by a functional of the parameters~$p$ and the sensor parameters~$z$ (the state variable~$u$ is implicitly given through the function~$g(p)$). These observables will be referred to as the \emph{model observations} $y \coloneqq h_{\text{obs}}(p,u,z)=h_{\text{obs}}(p,z)$. In the following, the term \emph{observables} will be used to refer to both experimental and model observations. 
In the example dealing with the launch of a spherical projectile, the experimental observables may consist in the altitude of the projectile, acquired by an optical apparatus or an altimeter, or the acceleration of the projectile obtained with an accelerometer. Whatever the quantity considered, it will be measured at a given discrete time~$z=t_\text{exp}$. The corresponding model observations are thus defined as $y(t_\text{exp}) \coloneqq u(t_{\text{exp}})$ or $y(t_\text{exp}) \coloneqq {\diff^2 u}/{\diff t^2}(t_{\text{exp}})$. 

\begin{figure}[tbh]
    \centering
    \footnotesize
    \def\figurescale{0.75} 
    \tikzset{external/export next=false}
    \pgfmathsetlengthmacro{\nodedistance}{1.5cm*\figurescale}
\pgfmathsetlengthmacro{\twidth}{4cm*\figurescale}
\begin{tikzpicture}[scale=\figurescale,node distance=\nodedistance]

\pgfdeclarelayer{bg}    
\pgfsetlayers{bg,main}  

\tikzstyle{io} = [trapezium, trapezium left angle=70, trapezium right angle=110,text width=\twidth,text centered,trapezium stretches body, draw=black, fill=blue!30]

\tikzstyle{process} = [rectangle,text width=\twidth,text centered, draw=black, fill=white]

\tikzstyle{decision} = [diamond,text width=\twidth,text centered, draw=black, fill=green!30]

\tikzstyle{parameter} = [trapezium, trapezium left angle=70, trapezium right angle=110,text width=0.75*\twidth,text centered,trapezium stretches body, draw=black, fill=white]

\tikzstyle{model_system} = [rectangle,text width=\twidth,text centered, draw=black,rounded corners=5pt, fill=white!90!black]

\tikzstyle{arrow} = [thick,-{Latex[length=2mm,width=2mm,round]}]

\node (origin) [coordinate] {};

\node (model parameter) [parameter] {\textbf{Model Parameters $\theta$}};

\node (obs functional) [process,below of=model parameter,yshift=-0.25*\nodedistance,xshift=-1.5*\nodedistance,text width=0.75*\twidth,draw=black,] {\textbf{Observation Functional $h_\text{obs}$}};
\node (qoi functional) [process,below of=model parameter,yshift=-0.25*\nodedistance,xshift=1.5*\nodedistance,text width=0.75*\twidth,draw=black] {\textbf{Quantity of Interest Functional $h_\text{qoi}$}};

\coordinate [left of=obs functional,xshift=-2.25*\nodedistance] (between param) {};

\node (control parameter) [parameter,above of=between param,yshift=-0.25*\nodedistance,text width=0.75*\twidth] {\textbf{Control Parameters $x$}};
\node (sensor parameter) [parameter,below of=between param,yshift=0.25*\nodedistance,text width=0.75*\twidth] {\textbf{Sensor Parameters $z$}};

\node (system) [model_system,left of=between param,xshift=-2.25*\nodedistance,text width=0.75*\twidth] {\textbf{System of Interest}};

\begin{pgfonlayer}{bg}
	\node (model) [model_system,fit=(model parameter)(obs functional)(qoi functional),inner sep=\nodedistance/3,label=above:\textbf{Model}] {};
\end{pgfonlayer}

\draw [arrow,name path=path1] (model parameter) -| (obs functional);
\draw [arrow] (model parameter) -| (qoi functional);

\draw [arrow] (control parameter) |- (system);
\draw [arrow] (between param) -- (obs functional);
\draw [thick] (sensor parameter) |- (between param);

\path [name path=path2] (control parameter) -- (control parameter -| model parameter) |- (qoi functional);

\path [name intersections={of=path1 and path2,by=inter1}];
\node (node_inter1) [] at (inter1){};
\draw [thick] (control parameter) -- (node_inter1.west);
\draw [arrow] (node_inter1.east) -- (control parameter -| model parameter) |- (qoi functional);

\node (exp obs) [process,below of=system,yshift=-0.75*\nodedistance,text width=0.85*\twidth] {\textbf{Experimental Observation $y_\text{exp}(x,z)$}};
\node (mod obs) [process,text width=0.85*\twidth] at (exp obs -| obs functional) {\textbf{Model Observation $y = h_\text{obs}(x,\theta,z)$}};
\node (qoi) [process,text width=0.85*\twidth] at (exp obs -| qoi functional) {\textbf{Quantity of Interest $q = h_\text{qoi}(x,\theta)$}};

\draw [arrow] (system) -- (exp obs);
\draw [arrow] (obs functional) -- (mod obs);
\draw [arrow] (qoi functional) -- (qoi);

\end{tikzpicture} 
    \caption{Diagram of the interactions between the various parameters and the system of interest as well as the model. It is important to observe that the model parameters~$\theta$ do not affect the system of interest, but only the observation and QoI functional~$h_\text{obs}$ and~$h_\text{qoi}$. For the sake of simplicity, the QoI functional~$h_\text{qoi}$ does not depend explicitly on the sensor parameters~$z$.}
    \label{fig:prediction_diagram}
\end{figure}
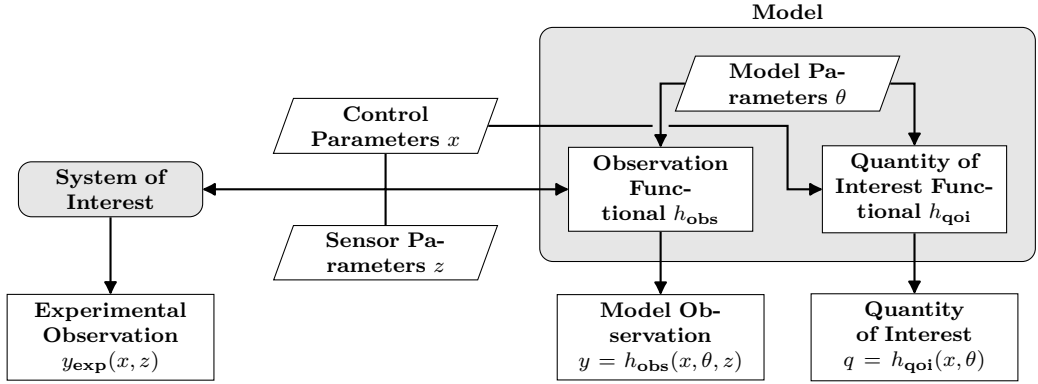

\emph{Quantities of interest} extend the notion of model observables, in the sense that they refer to quantities that can or cannot be experimentally observed. In other words, QoIs may not depend on sensor positions, although they could depend implicitly on the independent (spatial or temporal) variables through an average or a maximum. The QoIs will be denoted as $q \coloneqq h_{\text{qoi}}(p,u) = h_{\text{qoi}}(p)$, since they are evaluated for specific scenarios. We list below some examples of such possible QoIs:
\begin{itemize}[itemsep=0pt,topsep=2pt,parsep=0pt,leftmargin=25pt]
    \item Mean quantities over a subregion of $\Omega$ (e.g.\ mean displacement or stress in a solid);
    \item Minimal or maximal quantities and their location (e.g.\ maximal stress);
    \item Statistical quantities (e.g.\ probability of leakage of nuclear waste).
\end{itemize}
An important notion is that the QoIs provide an abstraction of the key features of the system of interest upon which a decision making process will be applied. Figure~\ref{fig:prediction_diagram} synthesizes how the observations and QoIs are obtained in terms of the various parameters. In the case of the projectile, an example of QoI could be the maximal altitude that it reaches over time, i.e.\ $q \coloneqq \max_{t} u(t)$. We note that the quantity is not directly observable since we would need to know beforehand the exact time at which the maximal altitude of the real projectile is reached.  

A precise description of the prediction problem can now be stated as follows: one wishes to predict a given QoI $h_\text{qoi}$ for a given prediction scenario $x_\text{pred}$. The definition of observable functionals $h_\text{obs}$ will play a crucial role in the calibration and validation problems in order to obtain accurate predictions. The accuracy of the model outputs (observables and QoI) also depends upon the aggregation of various sources of uncertainties and errors, as explained in the following subsection.

\subsection{Uncertainties and Modeling Errors} \label{ssec:uncertainty_errors}

Several approaches can be adopted to represent uncertainties in quantities, such as parameters, observables, or QoI, as outlined in~\cite{klir_uncertainty_2006, riedmaier_unified_2021}. We consider here the probabilistic approach, where an uncertain quantity is assimilated as a random variable~$A$ with probability distribution~$F_A$ and density function~$f_A$. The random variable~$A$ can be either discrete or continuous. We adopt the usual convention that a lowercase variable~$a$ indicates one realization of the random variable~$A$, noted in uppercase. In addition to random variables, we may also encounter random processes~$A=A(b)$, usually consisting of a family of random variables, each defined for an instance of~$b$.

Uncertainties are sometimes classified as \textit{aleatory} uncertainty and \textit{epistemic} uncertainty~\cite{roy_comprehensive_2011}. We will assume in this work that all uncertainties are aleatory. More precisely, this means that one is able to prescribe the distribution~$F_A$ of any random variable~$A$. In contrast, if the uncertainty were epistemic, we would not be able to characterize~$F_A$, see e.g.~\cite[Section 2]{roy_comprehensive_2011}. Epistemic uncertainty can thus be viewed here as \textit{uncertainty on the uncertainty}. We note that some authors define epistemic uncertainty as the uncertainty that could be eventually reduced if more information and data were available~\cite{hills_roll-up_2013, mullins_separation_2016}. However, such subtlety in the definition of epistemic uncertainty is not relevant to the current discussion and will be here passed over. Since we suppose that all uncertainties will be aleatory, it is reasonable to offer some guidance on how the distributions can be determined, as we do below.  

\subsubsection{Uncertainty in the Parameters}

As a first approximation, we suppose that the control parameters~$x$ and the sensor parameters~$z$ do not possess any uncertainty, an assumption that can be far from the truth depending on the nature of the prediction scenario and of the actual experimental apparatus. However, this hypothesis will simplify the definition of the optimal validation experiment as well as the overall exposition of the present work. Therefore, the only parameters that we will be considered uncertain here are the model parameters~$\theta$.


The quantification of the uncertainty in the model parameters~$\theta$ is a non-trivial task since this type of parameters arises from the various hypotheses of the model. One can determine the distribution~$F_\Theta$ by taking into account prior knowledge concerning each hypothesis. In the example of the projectile, we possess relatively extensive prior knowledge on the gravitational constant~$g$ and the air viscosity~$\mu$, so one can assign them a distribution with small variance.
However, it is often the case that we have a rather vague idea about the values of the model parameters. The probability distribution of each parameter need then be identified through calibration processes. This particular topic is the subject of an exhaustive literature. For the sake of completeness and to emphasize the difference between the problem of calibration and validation, we further detail the calibration of model parameters in Section~\ref{ssec:calibration}.

\subsubsection{Uncertainty in the Observables and QoIs}

Uncertainty in experimental observations may come from various sources and may depend on the scenarios~$x$ and sensor parameters~$z$. Experimental observations are thus assimilated as random processes~$Y_{\text{exp}}(x,z)$. Being able to characterize the uncertainty in the observables is crucial for the validation process, as further explained in Section~\ref{ssec:validation}. We list two possible, although not exhaustive, sources of uncertainty. The first one comes from the inherent and inevitable variability of the measurement apparatus. However, it is sometimes possible to reduce the amount of uncertainty in the observables and thus get experimental observations closer to the \textit{true} value by using a more precise apparatus. The second source of uncertainty comes from the inherent stochastic nature of the system of interest. For fixed control and sensor parameters~$x$ and~$z$, repetitions of the same experiment may yield different, yet accurate, values for the experimental observations. This may happen for instance if the phenomena of interest suffer from instabilities with respect to infinitesimal perturbations of the control parameters. The aggregation of these two sources of uncertainty allows one to determine the distribution~$F_{Y_{\text{exp}}(x,z)}$, but the task is not necessarily straightforward. We take the point of view that we can repeat a certain number of times the same experiments (in the case of non-destructive experiments) so that one can determine the empirical distribution~$F_{Y_{\text{exp}}(x,z)}$, which aggregates both sources of uncertainty.   

Uncertainty in the model observation~$Y$ and quantity of interest~$Q$ are obtained by propagating the uncertainty in the model parameters~$\Theta$ through their respective functional, that is, 
\begin{align}
    &Y(x,z) \coloneqq h_{\text{obs}}(x,\Theta,z)\\
    &Q(x) \coloneqq h_{\text{qoi}}(x,\Theta)
\end{align}
Computing the exact distributions~$F_{Y}$ and~$F_{Q}$ is only possible for very simple functionals and for a restricted family of distributions~$F_\Theta$ (e.g.\ Gaussian distributions). Most of the time, one needs to approximate the distributions~$F_{Y}$ and~$F_Q$ by sampling the distribution~$F_\Theta$ using methods such as the ubiquitous Monte-Carlo method, the Latin Hypercube method, or Quasi Monte-Carlo methods~\cite{sullivan_introduction_2015}. It is often enlightening to perform an uncertainty propagation prior to any validation step. If the prediction of the $Q$ possesses too much uncertainty, then the prediction may be useless, independently of whether the model is deemed valid or not. In this case, it may be necessary to refine the calibration of the model parameters~$\Theta$ in order to reduce their uncertainty, thereby potentially  reducing as well the uncertainty in the QoI. The use of sensitivity analysis methods~\cite{saltelli_global_2007,da_veiga_basics_2021} can help identify the model parameters~$\Theta$ contributing the most or least to the uncertainty in~$Y$ and~$Q$.
We summarize the various quantities introduced so far as well as the source of their uncertainty in Table~\ref{tab:quantity_summary}.

\begin{table}[t]
\caption{Summary of the parameters, observables, and QoI as well as their source of uncertainty}
\label{tab:quantity_summary}
\vspace{2pt}
\centering
\small
\begin{tabular}{@{}lcc@{}}
\toprule
 &
  \textbf{Definition} &
  \textbf{Uncertainty} \\ \midrule
\textbf{Control Parameters} $x$ &
  \begin{tabular}[c]{@{}c@{}}Parameters that one can control \\ to perform an experiment\end{tabular} &
  None \\ \addlinespace
\textbf{Model Parameters} $\Theta$ &
  \begin{tabular}[c]{@{}c@{}}Remaining parameters necessary\\ to describe a model\end{tabular} &
  \begin{tabular}[c]{@{}c@{}}Prior knowledge or\\ posterior density\end{tabular} \\ \addlinespace
\textbf{Sensor Parameters} $z$ &
  Spatio-temporal parameters &
  None \\ \midrule
\begin{tabular}[c]{@{}c@{}} \textbf{Experimental Observation}\\ $Y_{\text{exp}}(x,z)$\end{tabular} &
  \begin{tabular}[c]{@{}c@{}}Observations at control parameters $x$ \\ and sensor parameters $z$\end{tabular} &
  Empirical density \\ \addlinespace
\textbf{Model Observation} $Y(x,z)$ &
  $Y(x,z)\coloneqq h_{\text{obs}}(x,\Theta,z)$ &
  Uncertainty propagation \\ \addlinespace
\textbf{Quantity of Interest} $Q(x)$ &
  $Q(x)\coloneqq h_{\text{qoi}}(x,\Theta)$ &
  Uncertainty propagation \\ \bottomrule
\end{tabular}%
\end{table}

\subsubsection{Modeling Errors}
\label{ssec:model_error}

The authors adhere to the common idea that that no model, as sophisticated as it may be, would be able to capture the exact behavior of a system of interest, since a model is merely an abstraction of the latter and often relies on several hypotheses. Therefore any model comes with some modeling errors that should be estimated in one way or the other. Furthermore, the modeling error should account for the inevitable sources of uncertainty in both the experimental and model observations. One way to combine parameter uncertainties and modeling errors is in terms of the discrepancy~$E$:
\begin{align}
    E(x,z) = Y_{\text{exp}}(x,z) - Y(x,z) = Y_{\text{exp}}(x,z) - h_{\text{obs}}(x,\Theta,z). \label{eq:discrepancy_model}
\end{align} 
The discrepancy~$E(x,z)$ can be viewed as a random process in the sense that it consists of a random variable for each instance of the control parameters~$x$ and of the sensor parameters~$z$. This discrepancy~$E$ can be viewed as the aggregation of the experimental errors and the model inadequacy in the approach taken by Kennedy and O'Hagan~\cite{arendt_quantification_2012, kennedy_bayesian_2001}. An important property of the discrepancy~\eqref{eq:discrepancy_model} is that it is defined independently of the QoI. Other approaches can be adopted to quantify the modeling errors~\cite{hills_roll-up_2013, oliver_validating_2015}. For the sake of simplicity in the presentation, and since our focus is rather on the design of validation experiments, we omit to describe these methods here. We refer the interested reader to~\cite{mullins_survey_2016, riedmaier_unified_2021} for a summary of these methods.

\subsection{Calibration Process and Optimal Design of Calibration Experiments}
\label{ssec:calibration}

We provide here a brief overview of the calibration process and of the related problem dealing with the optimal design of calibration experiments. Although our proposed approach to perform optimal design of validation experiments does not directly involve a calibration process \textit{per se}, a brief discussion on the topic is deemed insightful in order to differentiate our methodology from the optimal design problem for calibration experiments.

The calibration of the model parameters~$\Theta$, a process usually referred to as model calibration or parameter identification, has as a main objective to provide a characterization of the probability distribution~$F_{\Theta}$, which encapsulates the uncertainty in the model parameters, using actual experimental observables~$y_\text{exp}$, as realizations of $Y_{\text{exp}}$. In other words, the goal is to update our knowledge about the model parameters. There exist several methods to perform a calibration, for example, the least squares approach (and its regularized version), the maximum likelihood estimation (MLE) approach, or the maximum a posteriori (MAP) approach. Under various hypothesis, one can show equivalences between these approaches~\cite{sivia_data_2006}. A common feature of these methods is that they provide a point estimator~$\hat{\theta}$ of the model parameters~$\theta$. These approaches present two drawbacks that are closely related. First, it is relatively difficult to quantify the uncertainty of~$\hat{\Theta}$ and only asymptotic results may be applicable~\cite[Section 16.4]{dasgupta_asymptotic_2008}. Second, if the model is not \emph{identifiable}, then these asymptotic results cannot be invoked and the point estimator~$\hat{\theta}$ may not be ``stable''. This identifiability condition is a restrictive condition to fulfil and especially tedious to verify~\cite{cole_parameter_2020}.


An alternative approach consists in updating the uncertainty of the model parameters with the posterior of a Bayesian analysis. As for the MLE and MAP approaches, the method requires defining a likelihood function~$L_{Y_\text{exp}=y_\text{exp},x_{\text{cal}},z_\text{cal}}(\theta)$, which describes the probability of obtaining the experimental observables~$Y_\text{exp} = y_\text{exp}$ given the control, sensor, and model parameters, $x_\text{cal}$,~$z_\text{cal}$, and~$\theta$, respectively. The definition of the likelihood function is not unique. We show below how to construct the likelihood function assuming an additive error
\begin{align}
    y_{\text{exp}}(x_{\text{cal}},z_\text{cal})  = y(x_{\text{cal}},z_\text{cal}) + e(x_{\text{cal}},z_\text{cal}) = h_{\text{obs}}(x_{\text{cal}},\theta,z_\text{cal}) + e(x_\text{cal},z_\text{cal}), \label{eq:additive_error}
\end{align} 
where~$y_{\text{exp}}(x_{\text{cal}},z_\text{cal})$ is a realization of the random variable~$Y_{\text{exp}}(x_{\text{cal}},z_\text{cal})$ and~$e(x_{\text{cal}},z_\text{cal})$ is a realization of the random variable~$E(x_{\text{cal}},z_\text{cal})$ that describes the error. We note that~$E$ is the same as in~\eqref{eq:discrepancy_model}. Since the error $E(x_{\text{cal}},z_\text{cal})$ is actually unknown, one can choose a reasonable approximation so that the likelihood be tractable. A typical choice is a Gaussian noise with zero mean and standard deviation~$\sigma>0$:
\begin{align}
    E \sim \mathrm{N}(0,\sigma^2).
\end{align} 
In other words, it is assumed that the noise is identical whatever the calibration scenario~$x_{\text{cal}}$ and sensor parameters~$z_\text{cal}$. The likelihood function resulting from this choice is defined as
\begin{align}
\label{eq:gaussian_likelihood}
L_{Y_\text{exp}=y_\text{exp},x_{\text{cal}},z_\text{cal}}(\theta) = \frac{1}{\sqrt{2\pi}\sigma}\exp\left[-\frac{\left(h_{\text{obs}}(x_{\text{cal}},\theta,z_\text{cal}) - y_{\text{exp}}(x_{\text{cal}},z_\text{cal})\right)^2}{2\sigma^2}\right]. 
\end{align}
Kennedy and O'Hagan~\cite{kennedy_bayesian_2001} propose to represent the discrepancy as the sum of a model error term (represented as a Gaussian process) and the experimental error. However, this approach is computationally involved and may lack identifiability since both the model parameters and the model errors need to be calibrated~\cite{arendt_quantification_2012}.

In addition to the likelihood, one must provide a prior density, denoted by~$\density_{\Theta_0}$, encoding the current knowledge that one possesses about the model parameters~$\Theta$. With the likelihood function and the prior density, Bayes' theorem allows one to obtain the posterior probability density function~$\density_{\Theta|Y_{\text{exp}}=y_\text{exp},x_\text{cal},z_\text{cal}}$ as
\begin{align}
\label{eq:bayesian_theorem}   \density_{\Theta|Y_{\text{exp}}=y_\text{exp},x_{\text{cal}},z_\text{cal}}(\theta) = \frac{L_{Y_\text{exp}=y_\text{exp},x_{\text{cal}},z_\text{cal}}(\theta)\density_{\Theta_0}(\theta)}{\int L_{Y_\text{exp}=y_\text{exp},x_{\text{cal}},z_\text{cal}}(\theta)\density_{\Theta_0}(\theta) \diff \theta}. 
\end{align}
The posterior density represents the updated knowledge about the model parameters~$\Theta$ given the experimental observations~$y_{\text{exp}}(x_{\text{cal}},z_\text{cal})$. The analytical expression of the posterior density is available only in a few cases, for example, when the model is linear with respect to the model parameters and one chooses a likelihood of the form~\eqref{eq:gaussian_likelihood} and a Gaussian prior. Otherwise, it is generally approximated using sampling methods such as the Markov Chain Monte-Carlo method or its variants~\cite{goodman_ensemble_2010,sullivan_introduction_2015}.

Since all calibration approaches require the use of experimental observations~$y_\text{exp}(x_\text{cal},z_\text{cal})$, the calibration scenario $x_\text{cal}$ and sensor parameters $z_\text{cal}$ should be chosen appropriately in order to better determine the uncertainty in the model parameters~$\Theta$. 
To the best of our knowledge, there essentially exists two methods for optimal design of calibration scenarios, one based on the Fisher information matrix and the other based on a Bayesian approach. 

\begin{enumerate}[itemsep=0pt,topsep=2pt,parsep=0pt,leftmargin=25pt]
\item 
The first method consists in optimizing some functional of the Fisher information matrix associated with the model~\cite{atkinson_optimal_2015,atkinson_optimum_2007}. This approach is widely used for polynomial models (linear with respect to the model parameters and possibly non-linear with respect to the control and sensor parameters) since the Fisher information matrix depends only upon the control and sensor parameters~\cite{atkinson_optimal_2015}. Different definitions of the Fisher information matrix functional lead to different optimal calibration experiments. For instance, the D-optimal design minimizes the variance of the estimator of the model parameters~$\theta$ and the G-optimal design minimizes the maximum of the variance in the prediction. The D-optimal problem for a linear model reads
\begin{align}
    (x_\text{cal}, z_\text{cal}) = \argmax_{(x,z) \, \in \, \mathcal{X}_\text{lab} \times \mathcal{Z}_\text{lab} } \, \abs*{\mathcal{I}(x,z)},
\end{align}
where $\mathcal{I}$ denotes the Fisher information matrix.
Equivalence theorems show that these different optimal designs are in fact related. For a thorough discussion of this approach and some extensions to non-linear models, we refer the interested reader to the book of Atkinson et al.~\cite{atkinson_optimum_2007}.

\item
The second method is based on the Bayesian approach~\cite{beck_fast_2018,long_fast_2013,ryan_review_2016} and focuses on minimizing some functional of the posterior~$f_{\Theta|Y=y_{exp},x_\text{cal},z_\text{cal}}$ (or sometimes the likelihood function) over the possible observations~$Y$ and the possible values of the model parameters~$\Theta$. Definitions of the functional akin to the one employed with the Fisher information matrix give rise to similar optimal designs. For example, we can seek the control parameters~$x_\text{cal}$ and sensor parameters~$z_\text{cal}$ that minimize the determinant of the covariance of the posterior
\begin{align}
    (x_\text{cal}, z_\text{cal}) = \argmax_{(x,z) \, \in \, \mathcal{X}_\text{lab} \times \mathcal{Z}_\text{lab} } \, \iint \big( \det(\Cov(\Theta|Y=y_{exp},x,z)) \big)^{-1} \, \diff F_{Y_\text{exp}}\, \diff F_{\Theta_0}.
\end{align}
We remark that solving these types of optimization problems can be very challenging because it usually requires a nested integration. Several simplifications, such as a Laplace approximation of the posterior, can make these optimization problems tractable~\cite{beck_fast_2018,long_fast_2013}.   
\end{enumerate}

A very important feature of the two approaches is that the determination of the optimal calibration experiment does not rely on any experimental data, as it is an a priori analysis. We will retain this feature in our methodology to design optimal validation scenarios. The whole calibration process is summarized in Algorithm~\ref{alg:calibration_process}.

\IncMargin{15pt}
\begin{algorithm}[tbh]
    \SetAlgoLined
    \vspace{5pt}
    \KwIn{Model~$(r,p,u)$, choice of observation functional~$h_{obs}$, choice of the error~$E(x,z)$, prior density~$\density_{\Theta_0}$, constrained sets~$\mathcal{X}_{\text{lab}}$ and~$\mathcal{Z}_\text{lab}$}
    Computation of the calibration scenario $x_{\text{cal}}$ and sensor parameters~$z_\text{cal}$ from optimal design of calibration experiments\;
    Realization of the experiment to obtain~$y_{\text{exp}}(x_\text{cal},z_\text{cal})$\;
    Computation of the posterior density~$\density_{\Theta|Y_{\text{exp}}=y_\text{exp},x_{\text{cal}},z_\text{cal}}$~\eqref{eq:bayesian_theorem} or approximation via sampling methods (such as MCMC)\;
    \KwOut{Posterior density~$\density_{\Theta|Y_{\text{exp}}=y_\text{exp},x_{\text{cal}},z_\text{cal}}$}
    \caption{Calibration Process}\label{alg:calibration_process}
\end{algorithm}
\DecMargin{15pt}

\subsection{Validation Process}
\label{ssec:validation}

Everything is now in place to discuss the problem of interest, which is the validation of a model. Again, the goal is to assess the accuracy in the prediction of a quantity of interest~$Q=h_{\text{qoi}}(x_{\text{pred}},\Theta)$ at the prediction scenario~$x_{\text{pred}}$. We reiterate that the discrepancy~$E$ given in~\eqref{eq:discrepancy_model} is a useful way to encode the accuracy of our model observations. Both experimental and model observations should be performed at a specific scenario, the so-called validation scenario~$x_{\text{val}}$, and for given sensor parameters~$z_\text{val}$. For now, let us suppose that we have determined this validation scenario~$x_{\text{val}}$, the type of observable (i.e.\ the choice of the functional~$h_{\text{obs}}$) and the value of the sensor parameters~$z_\text{val}$. Multiple approaches to validate a model have been proposed, as mentioned in the Introduction. Several of these validation processes require the choice of a validation metric. For example, Roy and Oberkampf~\cite{roy_comprehensive_2011} utilize the area validation metric introduced by Ferson et al.~\cite{ferson_model_2008}
\begin{align}
\label{eq:area_validation_metric}
    d(F_{Y},F_{Y_\text{exp}}) = \int_{-\infty}^\infty \abs*{F_Y(s) - F_{Y_\text{exp}}(s)} \, \diff s, 
\end{align}
where~$F_Y$ and~$F_{Y_\text{exp}}$ are the model and experimental observation distributions, respectively. They provide several rationales for this choice of validation metric and provide several ways to employ it, especially in the presence of different validation scenarios. Another validation metric employs the discrepancy~$E$ with a tolerance~$\varepsilon$ on the error
\begin{align}
\label{eq:validation_metric}
    \gamma = F_{\abs*{E}}(\varepsilon) = \Prob(\abs*{E} < \varepsilon) = \Prob(\abs{Y_{\text{exp}}(x_{\text{val}},z) - h_{\text{obs}}(x_{\text{val}},\Theta,z)} < \varepsilon). 
\end{align}
If the value of~$\gamma$ is above a certain threshold~$\eta$, then the model is deemed valid. This particular measure was introduced by Rebba and Mahadevan~\cite{rebba_computational_2008} and coined \textit{model reliability metric}. In their paper, the authors compare the model reliability metric with point and interval hypothesis testing under both the frequentist and Bayesian perspectives. They found that adopting the hypothesis testing approach may lead to some inconsistencies as to whether the model is deemed valid or not, whereas the model reliability metric dos not present these inconsistencies. Li and Mahadevan~\cite{li_role_2016} expand this model reliability metric to take into account multivariate outputs. Mullins et al.~\cite{mullins_separation_2016} further investigate the use of the model reliability metric for validation. In this paper, the authors argue that the use of the aforementioned area validation metric can be misleading. Indeed, increasing the uncertainty of the experimental observation so that~$F_{Y_\text{exp}} \approx F_{Y}$ leads to a smaller area validation metric (a valid model), whereas one should have less confidence in the validity of our model, since the experimental data are of lesser quality. For these reasons, we will select the reliability metric~\eqref{eq:validation_metric} as the validation metric for the remainder of the paper.  

To compute the probability~$\gamma$, one needs to know the distribution of the discrepancy~$E$. We assume here that the random variables~$Y_{\text{exp}}(x_\text{cal},z)$ and~$\Theta$ are independent. This assumption is reasonable since the first quantity relates to the experimental observations while the second quantity is related to the various hypotheses on the model. We can then estimate~$\gamma=F_{\abs*{E}}(\varepsilon)$ with any of the previously mentioned sampling methods by performing the sampling of the empirical distributions of~$Y_{\text{exp}}(x_\text{cal},z)$ and of the model parameters~$\Theta$ independently.

\subsubsection{Validation Scenario}

The choice of the validation scenario~$x_{\text{val}}$, the type of observable, and the value of the sensor parameters~$z_\text{val}$ have a significant impact on the actual outcome of the validation process, as it will be illustrated in Section~\ref{ssec:example1}. It is seldom the case that we can replicate in a controlled environment the conditions under which we would like to perform the prediction, wherein the prediction and validation scenario are the same. If it is not the case, then we must come up with an \textit{informative} and \textit{relevant} validation scenario with respect to the prediction scenario. Moreover, if the QoI is not observable, then we need to determine which type of observables and which sensor parameters~$z$ should be employed for the validation. These are the main issues addressed in this research work, which will be further explored in Section~\ref{sec:optimal_design_validation}.

\subsubsection{Validation Workflow}

A complementary objective of our work is to eliminate as much as possible arbitrary decisions in the validation process. Indeed, by providing a way to determine the validation scenario~$x_{\text{val}}$, the type of observables, and the sensor parameters~$z_\text{val}$, we are closer to \textit{automating} the validation process. Our perspective of what a validation workflow should look like is summarized by Algorithm~\ref{alg:validation_process}.

\IncMargin{15pt}
\begin{algorithm}[tbh]
    \SetAlgoLined
    \vspace{5pt}
    \KwIn{Model~$(r,p,u)$, quantity of interest~$h_{\text{qoi}}$, prediction scenario~$x_{\text{pred}}$, choice of observation functional~$h_{obs}$, choice of the error~$E(x,z)$, prior density~$\density_{\Theta_0}$, constrained sets~$\mathcal{X}_{\text{lab}}$ and~$\mathcal{Z}_\text{lab}$, error tolerance~$\varepsilon$, threshold~$\eta$}
    Set~$\density_\Theta = \density_{\Theta_0}$ \;
    \If{Calibration experiment available}{
    Calibration of the model parameter~$\Theta$ (Algorithm~\ref{alg:calibration_process})\;
    Set~$\density_\Theta = \density_{\Theta|Y_{\text{exp}}=y_\text{exp},x_{\text{cal}},z_\text{cal}}$ \;
    }
    Computation of~$Q=h_{\text{qoi}}(x_{\text{pred}},\Theta)$\;
    \eIf{Uncertainty of~$Q=h_{\mathrm{qoi}}(x_{\mathrm{pred}},\Theta)$ is too large}{Exit validation\;}{Proceed to validation\;}
    Computation of the validation scenario~$x_{\text{val}}$, sensor parameters~$z_\text{val}$,  and type of observation functional~$h_{\text{obs,val}}$ (Algorithm~\ref{alg:optimal_design_validation})\;
    Realization of the experiment to obtain~$y_{\text{exp}}(x_\text{val},z_\text{val})$\;
    Computation of the validation metric~$\gamma$~\eqref{eq:validation_metric}\;
    \eIf{$\gamma\geq \eta$}{Model is not invalidated}{Model is  invalidated}
    \caption{Validation Process}\label{alg:validation_process}
\end{algorithm}
\DecMargin{15pt}

The first step of the validation process consists in defining the prior density~$\density_{\Theta_0}$ of the model parameters~$\Theta$ using prior knowledge or via a calibration process, such as the one described in Algorithm~\ref{alg:calibration_process}. The second step consists in a \textit{sanity check}, that of checking whether the prediction of $Q=h_{\text{qoi}}(x_{\text{pred}},\Theta)$ has too much uncertainty given the uncertainty of the model parameters~$\Theta$. If it does, then it is irrelevant to continue the validation process since the prediction would not be useful, independently of whether or not the model is valid. If the uncertainty in the QoI is deemed acceptable, then one needs to design an appropriate validation experiment. One would then carry out the validation experiment and estimate the validation metric to verify that its prediction indeed reflects the reality. 

As a final remark, we do not address in this work the propagation of the modeling errors to the QoI at the prediction scenario, nor the decision making about the model validity. Our objective is mainly to search for the best validation experiments so that one can better assess the capability of the model to predict the QoI. That said, we conjecture that our approach for designing optimal validation experiments, as described in the next section, complements the validation workflow presented in Algorithm~\ref{alg:validation_process} to provide useful assessments of the model validity.

\section{Optimal Design of Validation Experiments} \label{sec:optimal_design_validation}

A validation scenario to obtain validation data must be \textit{representative} of the prediction scenario, since the objective is to assert whether or not the model is valid for predictive purposes, more specifically, for the prediction of the quantity of interest~$Q=h_\text{qoi}(x_\text{pred},\Theta)$). By \textit{representative}, we mean that the various hypotheses on the model should be similarly satisfied in both the prediction and validation scenarios. In many applications, it is customary to design the validation experiment based on dimensionless numbers. In fluid mechanics for instance, the Reynolds number is often used as a criterion to select validation conditions that reflect the prediction setting. However, this simple choice does not necessarily provide a complete framework for the design of validation experiments. First, a model usually involves more parameters than units and several dimensionless numbers need to be considered simultaneously. Moreover, the relative influence of each of the dimensionless numbers cannot be solely provided by dimensional analysis. It is in fact important to quantify the influence of a dimensionless number on the QoI for the following two reasons: 
\begin{enumerate}[itemsep=0pt,topsep=2pt,parsep=0pt,leftmargin=25pt]
    \item the uncertainty in the model parameters~$\Theta$ is propagated to the dimensionless parameters;
    \item the dimensionless numbers in the validation and prediction settings rarely match exactly.
\end{enumerate}
Although dimensional analysis is a useful tool to better understand a model, it seems insufficient and somewhat ill-defined for validation purposes. Our view is that the behavior of the model with respect to the parameters and their uncertainty should be as similar as possible in the validation and prediction scenarios. In other words, an analysis of the parameter sensitivity should guide the optimal design of validation experiments.

\subsection{Sensitivity Analysis}
\label{ssec:sensitivity_analysis}

Sensitivity analysis is a vast subject with multiple objectives and methodologies, see e.g.~\cite{razavi_future_2021,saltelli_global_2007,da_veiga_basics_2021} for thorough overviews on the  topic. Examples of problems addressed by sensitivity analysis are:
\begin{itemize}[itemsep=0pt,topsep=2pt,parsep=0pt,leftmargin=25pt]
    \item Characterization of the relationship between the parameters and the model outputs.
    \item Identification of non-influential parameters, a topic often referred in the literature to as \textit{factor fixing}.
    \item Identification of the parameters that influence the most the model outputs, usually referred to as \textit{factor prioritization}. One objective here is to further reduce the uncertainty affecting these parameters.
\end{itemize} 

Sensitivity methods are usually classified as either local or global. On one hand, local methods compute the sensitivity (e.g.\ derivatives) around a specific point in the parameter space and do not take into account the uncertainty affecting the parameters. On the other hand, global methods incorporate the uncertainty in the parameters. The Sobol'-based sensitivity analysis is a global approach that computes the so-called Sobol' sensitivity indices, which quantitifies how much of the variance in a model output can be attributed by a single parameter or a subset of parameters~\cite{sobol_global_2001}. 

We consider in our approach that the model parameters~$\Theta$ are uncertain so that we need to employ a global sensitivity analysis method. Moreover, our goal being to assess the influence of the various parameters on the model outputs, either the observables or the QoI (the first objective aforementioned), we seek a description of the response surface of the observation and QoI functionals with respect to the parameters. Our objective is thus different from that of allocating the variance of~$Y=h_\text{obs}(x,\Theta,z)$ or~$Q=h_\text{qoi}(x,\Theta)$ to parameters (or subsets of parameters) such as in the Sobol' approach.


\subsection{Active Subspace Method}
\label{ssec:active_subspace}

One sensitivity method that allows the characterization of a response surface while taking into account the uncertainty of the parameters is the Active Subspace method~\cite{constantine_active_2014,constantine_active_2015}. The method computes the gradient of a model functional~$h$ with respect to the control and model parameters to quantify their influence. The expectation of the outer product of the gradient with respect to the distribution~$F_\Theta$ is performed to assemble the so-called \emph{influence matrix}~$M_h$
\begin{equation}
\label{eq:matrix_M}
\begin{aligned}
    M_h(x,z) &= \Exp\left(\nabla_{(x,\theta)} h(x,\Theta,z) \, \nabla_{(x,\theta)} h(x,\Theta,z)^T\right) \\
    &= \Cov\left(\nabla_{(x,\theta)} h(x,\Theta,z)\right) + \Exp\left(\nabla_{(x,\theta)} h(x,\Theta,z)\right) \Exp\left(\nabla_{(x,\theta)} h(x,\Theta,z)\right)^T .
\end{aligned}
\end{equation}
One proceeds with the eigenvalue decomposition of~$M_h(x,z)$ to identify the principle directions influencing the outputs of the model. This influence matrix is characterized by large variations in the gradient and/or large expected gradients. From now on, the influence matrix~$M_h$ will constitute our quantitative description of the surface response of the functional~$h$.

Given two influence matrices~$M_1$ and $M_2$ characterized by different functionals~$h$, different scenarios~$x$, and different sensor parameters~$z$, we define the distance between the two influence matrices as
\begin{gather}
\label{eq:distance_M}
    d(M_1,M_2) = \norm*{M_1 - M_2}_2, 
\end{gather}
where $\norm*{M}_2$ denotes the spectral norm of $M$ (other norms could have been considered as well). The distance essentially measures the error between an influence matrix~$M_1$ and a target influence matrix~$M_2$. We note that the use of the Active Subspace method presumes enough regularity of the observations and QoI functionals.

Having defined a quantification of the response surface of a functional as well as a means for comparison, we turn our attention to the problem at hand, that of defining a validation experiment specifically tailored toward the prediction of the QoI at a given prediction scenario.

\subsection{Optimal Choice of the Validation Scenario}
\label{ssec:design_validation_scenario}

We describe the set of controlled experiments as the set of scenarios~$\mathcal{X}_\text{lab}\subseteq \mathcal{X}$ for which one is able to make a laboratory experiment to interrogate the system of interest. We recall that the prediction scenario $x_\text{pred}$ may or may not belong to $\mathcal{X}_\text{lab}$. We thus formulate two requirements regarding an optimal validation scenario:

\begin{enumerate}
[itemsep=0pt,topsep=2pt,parsep=0pt,leftmargin=25pt]
    \item If~$x_\text{pred} \in \mathcal{X}_\text{lab}$, then the validation scenario should be given by~$x_\text{val}=x_\text{pred}$.
    \item If the QoI functional~$h_\text{qoi}$ is in fact an observable~$h_\text{obs}$, then the observable and the sensor parameters employed for the validation should correspond to those of the QoI.
\end{enumerate}

These two requirements will guide the definition of the two optimal design problems that will be set up to specify the goal-oriented validation experiment. The first step in specifying the validation experiment consists in determining the validation scenario~$x_\text{val}$. We recall here that the QoI functional~$h_\text{qoi}$ is only a function of the control parameters~$x$ and of the model parameters~$\Theta$. Moreover, the model parameters~$\Theta$ are the same for the validation and prediction scenarios, so we have two random quantities of interest~$Q_\text{val} = h_\text{qoi}(x_\text{val},\Theta)$ and~$Q_\text{pred} = h_\text{qoi}(x_\text{pred},\Theta)$. We seek the validation scenario~$x_\text{val}$ that minimizes the distance~\eqref{eq:distance_M}, with the constraint that the scenario be realizable in a controlled environment  
\begin{align}
\label{eq:obj_function_control}
    x_\text{val} \in \argmin_{x\in\mathcal{X}_\text{lab}} d(M_{h_\text{qoi}}(x),M_{h_\text{qoi}}(x_\text{pred})).
\end{align}
In other words, we look for the validation scenario such that the response surface of~$Q_\text{val}$ (encoded by the influence matrix~$M_{h_\text{qoi}}(x_\text{val})$) resembles the most the response surface of~$Q_\text{pred}$ for the prediction scenario. This is another way of saying that the various hypotheses and simplifications leading to the model have comparable impact in the validation and predictive settings. We observe that if~$x_\text{pred}\in\mathcal{X}_\text{lab}$, then a global minimum with zero value is attained at~$x_\text{pred}=x_\text{val}$, since the objective function in~\eqref{eq:obj_function_control} is non-negative. However, the global minimum may not be unique. Indeed, a linear functional~$h_\text{qoi}$ with respect to the parameters features the same influence matrix~$M_{h_\text{qoi}}(x)$ whatever the value of these parameters. Hence, the first requirement stated above is partially fulfilled, since the optimization problem~\eqref{eq:obj_function_control} may not possess a unique minimum. 

\subsection{Optimal Choice of Observables and Sensor Parameters}
\label{ssec:design_observable_sensor}

The goal in this section is to define the problem whose solution will provide an observation functional~$h_\text{obs}$ and sensor parameters~$z$ in order to fully characterize the optimal validation experiment.    
We recall that QoIs are not necessarily observable either in the prediction or the validation scenarios. We nevertheless want to validate the model for the prediction of the QoI. With an optimal prediction scenario obtained by solving~\eqref{eq:obj_function_control}, we now need to determine the functional~$h_\text{obs}$ and sensor parameters~$z$ that best mimic the influence matrix of~$Q_\text{val}=h_\text{qoi}(x_\text{val},\Theta)$. Again, the rationale is that the influence matrix reflects the impact of the various hypotheses and assumptions of the model. The result of the validation process obtained with this particular functional~$h_\text{obs}$ and sensor parameters~$z$ could then be reasonably extrapolated to the QoI. Since the choices of the observation functional and of the sensor parameters~$z$ are constrained by the available experimental capabilities, we introduce here the set $\mathcal{H}_\text{lab}$ of possible observation functionals and the set $\mathcal{Z}_\text{lab}$ of possible sensor parameters. The optimal design problem to be solved is then given by
\begin{align}
\label{eq:obj_function_sensor}
    (h_\text{obs,val},z_\text{val}) \in \argmin_{(h_\text{obs},z) \in \mathcal{H}_\text{lab} \times \mathcal{Z}_\text{lab}} d\bigg(\frac{M_{h_\text{obs}}(x_\text{val},z)}{\Tr(M_{h_\text{obs}}(x_\text{val},z))},\frac{M_{h_\text{qoi}}(x_\text{val})}{\Tr(M_{h_\text{qoi}}(x_\text{val}))}\bigg).  
\end{align}
The normalization of the influence matrices for the observational and QoI functional is required since both functionals may represent different physical quantities with different scaling and units. The validation experiment tailored toward the prediction of the QoI is now fully specified. It consists in interrogating the system of interest for the experimental observation corresponding to the observation functional~$h_\text{obs,val}$ at the sensor parameters~$z_\text{val}$ under the scenario~$x_\text{val}$. The whole process to search for the optimal experiment is described in Algorithm~\ref{alg:optimal_design_validation}.

\IncMargin{15pt}
\begin{algorithm}[t]
    \SetAlgoLined
    \vspace{5pt}
    \KwIn{Model~$(r,p,u)$, quantity of interest~$h_{\text{qoi}}$, prediction scenario~$x_{\text{pred}}$, model parameter~$\Theta$, set of controlled scenarios~$\mathcal{X}_\text{lab}$, set of observation functionals~$\mathcal{H}_\text{lab}$, set of controlled sensor parameters~$\mathcal{Z}_\text{lab}$}
    Computation of the validation scenario~$x_\text{val}$ by solving~\eqref{eq:obj_function_control}\;
    Computation of the validation observation functional~$h_\text{obs,val}$ and validation sensor parameter~$z_\text{val}$ by solving~\eqref{eq:obj_function_sensor}\;
    \KwOut{Validation scenario~$x_\text{val}$, validation observation functional~$h_\text{obs,val}$, validation sensor parameters~$z_\text{val}$}
    \caption{Optimal Design of Validation Experiment}\label{alg:optimal_design_validation}
\end{algorithm}
\DecMargin{15pt}

The values of the objective functionals of Problems~\eqref{eq:obj_function_control} and~\eqref{eq:obj_function_sensor} both provide useful information regarding the quality of the validation experiment. If both objective functionals are small, then the influence matrix of the observables~$h_\text{obs}$ at the validation scenario mimics well the influence matrix of the QoI at the prediction scenario. The validation experiment is thus related to the predictive setting. If the objective functional of problem~\eqref{eq:obj_function_control} is large, then one may need to consider expanding the set of controlled scenarios~$\mathcal{X}_\text{lab}$. Also, if the objective functional of Problem~\eqref{eq:obj_function_sensor} is large, then one may seek other types of experimental measurements. In case of one or both situations, one must exercise caution in extrapolating the result of the validation process to the prediction scenario.

The solution of the optimization problems~\eqref{eq:obj_function_control} and~\eqref{eq:obj_function_sensor} is performed with the solver NOMAD~\cite{audet_algorithm_2022} and is further detailed in~\ref{ann:optimization}.

\section{Numerical Examples}
\label{sec:examples}

We illustrate our methodology on two examples, the first one consisting in the projectile problem introduced in Section~\ref{ssec:abstract_model} and the second one consisting in a pollutant transport problem. The first problem is used to illustrate the entire validation process and to emphasize the importance of the design of the validation experiment. In particular, we will provide an example of a poorly chosen validation experiment that leads to a false positive, that is, the model is not deemed invalid when it is in fact invalid. By contrast, we will show that the application of our methodology to design an optimal validation experiment produces a true negative. The second problem will show the importance of the definition of the QoI on the validation experiment.

\subsection{The Projectile Problem}
\label{ssec:example1}

The first example considers the system of interest described in Section~\ref{ssec:abstract_model}, where we seek the maximum altitude reached by a spherical projectile when launched vertically. The model is given by the linear ordinary differential equation~\eqref{eq:model_projectile} for which one can analytically compute the value of the QoI as
\begin{align}
\label{eq:qoi_projectile}
    q(x,\theta) = \max_t u(t) = u_0+ \frac{mv_0}{3\pi \exp(\mu) \ell} - \frac{m^2g}{(3\pi \exp(\mu) \ell)^2} \ln\left(1+\frac{3\pi \exp(\mu) \ell v_0 }{mg}\right).
\end{align}
We recall that the control parameters are given here by $x=(m,\ell,u_0,v_0)$ while the model parameters are $\theta=(g,\mu)$. Since the proposed QoI is non-linear with respect to some of the control parameters, the associated influence matrix~\eqref{eq:matrix_M} depends on the scenario~$x$. Moreover, we set the model parameters to~$\Theta = (G,U) \sim (\mathrm{N}(9.81,0.01^2),\mathrm{N}(-5\ln(10),0.5^2))$, where $\mathrm{N}$ denotes normal distributions. The values were chosen based on the fact that we know relatively well the value of the gravitational constant~$g$ but not as well the air viscosity~$\mu$.

\subsubsection{Optimal Validation Scenario}
\label{ssec:optimal_val_scenario_projectile}

To illustrate the possibility that a poorly designed validation experiment may yield a false positive, we need to consider a specific prediction scenario~$x_\text{pred}$ and a specific controlled environment~$\mathcal{X}_\text{lab}$ that describes the possible validation scenarios~$x_\text{val}$. The prediction scenario and a tentative validation scenario, as well as the controlled environment, are described in Table~\ref{tab:scenario_false_positive}.

\begin{table}[t]
\caption{Prediction scenario~$x_\text{pred}$, controlled environment~$\mathcal{X}_\text{lab}$, and tentative validation scenario~$x_\text{val,tent}$}
\label{tab:scenario_false_positive}
\vspace{2pt}
\begin{tabular}{@{}ccccc@{}}
\toprule
 &
  \textbf{Mass}~$m$ &
  \textbf{Diameter}~$\ell$ &
  \textbf{Initial Position}~$u_0$ &
  \textbf{Initial Velocity}~$v_0$ \\ \midrule
\begin{tabular}[c]{@{}c@{}}
\textbf{Prediction}\\ 
\textbf{Scenario}~$x_\text{pred}$
\end{tabular} &
  0.05 &
  0.01 &
  1 &
  100 \\ \midrule
\begin{tabular}[c]{@{}c@{}}
\textbf{Controlled} \\ 
\textbf{Environment}~$\mathcal{X}_\text{lab}$
\end{tabular} & 
  [1,5] & 
  [0.05,0.1] & 
  [0,2] & 
  [10,120] \\ \midrule
\begin{tabular}[c]{@{}c@{}}
\textbf{Tentative Validation}\\ 
\textbf{Scenario}~$x_\text{val,tent}$
\end{tabular}&
  2.5 &
  0.05 &
  1 &
  20\\ \bottomrule
\end{tabular}
\end{table}

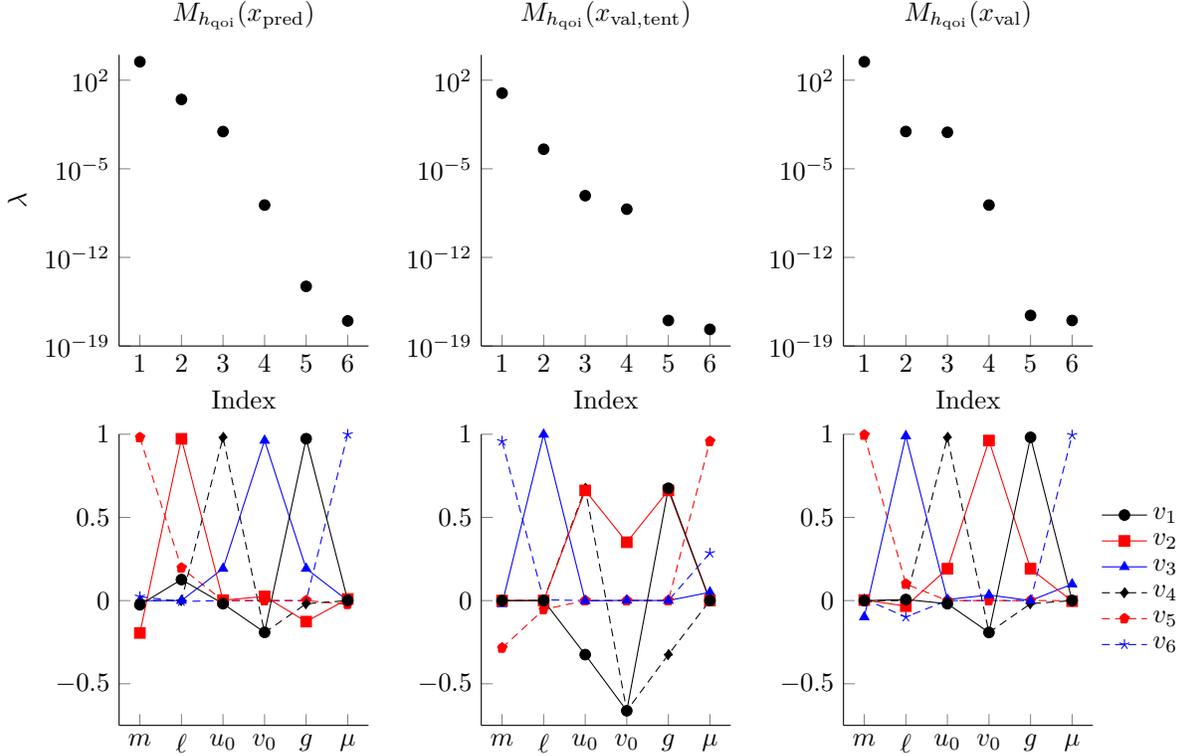
\begin{figure}[t]
    \centering
    \setlength\figureheight{0.25\linewidth}
    \setlength\figurewidth{0.225\linewidth}
    \setlength{\tabcolsep}{18pt}
    \begin{tabular}{lcr}
        \tikzset{external/export next=false}
        \begin{tikzpicture}[baseline,trim axis left, trim axis right]
\begin{axis}[
    width=0.951\figurewidth,
    height=\figureheight,
    at={(0\figurewidth,0\figureheight)},
    scale only axis,
    xlabel=Index,
    xtick={1,2,3,4,5,6},
    ymin=1e-19,
    ymax=1e4,
    ylabel=$\lambda$,
    ymode = log,
    axis x line*=bottom,
    axis y line*=left,
    legend style={at={(0,1)}, anchor=north west, draw=none,fill=none},
    title={$M_{h_\text{qoi}}(x_\text{pred})$},
    ]
    \pgfplotstableread{./Results/projectile/eigenvalue.txt}\loadedtable
    
    \addplot [black,only marks] table [x=index,y=lambda_pred] {\loadedtable};

\end{axis}
\end{tikzpicture} &
        \tikzset{external/export next=false} 
        \begin{tikzpicture}[baseline,trim axis left, trim axis right]
\begin{axis}[
    width=0.951\figurewidth,
    height=\figureheight,
    at={(0\figurewidth,0\figureheight)},
    scale only axis,
    xlabel=Index,
    xtick={1,2,3,4,5,6},
    ymin=1e-19,
    ymax=1e4,
    ymode = log,
    axis x line*=bottom,
    axis y line*=left,
    legend style={at={(0,1)}, anchor=north west, draw=none,fill=none},
    title={$M_{h_\text{qoi}}(x_\text{val,tent})$},
    ]
    \pgfplotstableread{./Results/projectile/eigenvalue.txt}\loadedtable
    
    \addplot [black,only marks] table [x=index,y=lambda_tent] {\loadedtable};

\end{axis}
\end{tikzpicture} &
        \tikzset{external/export next=false}
        \begin{tikzpicture}[baseline,trim axis left, trim axis right]
\begin{axis}[
    width=0.951\figurewidth,
    height=\figureheight,
    at={(0\figurewidth,0\figureheight)},
    scale only axis,
    xlabel=Index,
    xtick={1,2,3,4,5,6},
    ymin=1e-19,
    ymax=1e4,
    ymode = log,
    axis x line*=bottom,
    axis y line*=left,
    legend style={at={(0,1)}, anchor=north west, draw=none,fill=none},
    title={$M_{h_\text{qoi}}(x_\text{val})$},
    ]
    \pgfplotstableread{./Results/projectile/eigenvalue.txt}\loadedtable
    
    \addplot [black,only marks] table [x=index,y=lambda_optim] {\loadedtable};

\end{axis}
\end{tikzpicture} \\
        \tikzset{external/export next=false}
        \begin{tikzpicture}[baseline,trim axis left, trim axis right]
\begin{axis}[
    width=0.951\figurewidth,
    height=\figureheight,
    at={(0\figurewidth,0\figureheight)},
    scale only axis,
    xtick={1,2,3,4,5,6},
    xticklabels={$m$,$\ell$,$u_0$,$v_0$,$g$,$\mu$},
    ymin=-0.75,
    ymax=1,
    yminorgrids=true,
    axis x line*=bottom,
    axis y line*=left,
    legend style={at={(1,0.5)}, anchor=west, draw=none,fill=none},
    cycle list name=eigenvectors,
    ]
    \pgfplotstableread{./Results/projectile/eigenvector_pred.txt}\loadedtable
    
    \addplot  table [x=index,y=v6] {\loadedtable};
    \addplot+ table [x=index,y=v5] {\loadedtable};
    \addplot+ table [x=index,y=v4] {\loadedtable};
    \addplot+ table [x=index,y=v3] {\loadedtable};
    \addplot+ table [x=index,y=v2] {\loadedtable};
    \addplot+ table [x=index,y=v1] {\loadedtable};
    
\end{axis}
\end{tikzpicture} & 
        \tikzset{external/export next=false}
        \begin{tikzpicture}[baseline,trim axis left, trim axis right]
\begin{axis}[
    width=0.951\figurewidth,
    height=\figureheight,
    at={(0\figurewidth,0\figureheight)},
    scale only axis,
    xtick={1,2,3,4,5,6},
    xticklabels={$m$,$\ell$,$u_0$,$v_0$,$g$,$\mu$},
    ymin=-0.75,
    ymax=1,
    yminorgrids=true,
    axis x line*=bottom,
    axis y line*=left,
    legend style={at={(0,1)}, anchor=north west, draw=none,fill=none},
    cycle list name=eigenvectors,
    ]
    \pgfplotstableread{./Results/projectile/eigenvector_tentative.txt}\loadedtable
    
    \addplot  table [x=index,y=v6] {\loadedtable};
    \addplot+ table [x=index,y=v5] {\loadedtable};
    \addplot+ table [x=index,y=v4] {\loadedtable};
    \addplot+ table [x=index,y=v3] {\loadedtable};
    \addplot+ table [x=index,y=v2] {\loadedtable};
    \addplot+ table [x=index,y=v1] {\loadedtable};

\end{axis}
\end{tikzpicture} &
        \tikzset{external/export next=false}
        \begin{tikzpicture}[baseline,trim axis left, trim axis right]
\begin{axis}[
    width=0.951\figurewidth,
    height=\figureheight,
    at={(0\figurewidth,0\figureheight)},
    scale only axis,
    xtick={1,2,3,4,5,6},
    xticklabels={$m$,$\ell$,$u_0$,$v_0$,$g$,$\mu$},
    ymin=-0.75,
    ymax=1,
    yminorgrids=true,
    axis x line*=bottom,
    axis y line*=left,
    legend style={at={(1,0.5)}, anchor=west, draw=none,fill=none},
    reverse legend,
    cycle list name=eigenvectors,
    ]
    \pgfplotstableread{./Results/projectile/eigenvector_optim.txt}\loadedtable
    
    \addplot  table [x=index,y=v6] {\loadedtable};
    \addplot+ table [x=index,y=v5] {\loadedtable};
    \addplot+ table [x=index,y=v4] {\loadedtable};
    \addplot+ table [x=index,y=v3] {\loadedtable};
    \addplot+ table [x=index,y=v2] {\loadedtable};
    \addplot+ table [x=index,y=v1] {\loadedtable};
    
    \legend{$v_6$,$v_5$,$v_4$,$v_3$,$v_2$,$v_1$}

\end{axis}
\end{tikzpicture}
    \end{tabular}
    \caption{Eigenvalues (top row) and eigenvectors (bottom row) associated with the influence matrices of the QoI~\eqref{eq:qoi_projectile} for the prediction scenario~$x_\text{pred}$ (left), the tentative validation scenario~$x_\text{val,tent}$ (center), and the optimal validation scenario~$x_\text{val}$ (right).}
    \label{fig:comparison_si_qoi_false_positive}
\end{figure}

The tentative validation scenario is chosen such that the Reynolds number $\mathrm{Re}$ taken at the initial time,
\begin{equation}
\label{eq:Reynoldsnumber}
    \mathrm{Re}(t) = \frac{\rho \ell}{\exp(\mu)} \abs*{\frac{\diff u}{\diff t}(t)},
\end{equation}
where $\rho$ is the density of the ambient air,
matches the one of the prediction scenario~$x_\text{pred}$. The rationale echoes the discussion at the beginning of Section~\ref{sec:optimal_design_validation} and, in this regard, may be considered as a viable validation experiment. However, as shown in Figure~\ref{fig:comparison_si_qoi_false_positive}, the influence matrix of the QoI at~$x_\text{val,tent}$ is fairly different from the influence matrix of the QoI at~$x_\text{pred}$. The controlled environment~$\mathcal{X}_\text{lab}$ consists in the Cartesian product of the intervals for the parameters $x$, as defined in Table~\ref{tab:scenario_false_positive}. We now solve problem~\eqref{eq:obj_function_control} to obtain the optimal validation scenario for this predictive setting and obtain~$x_\text{val} = (1.000,0.100,0.705,100.446)$. Figure~\ref{fig:comparison_si_qoi_false_positive} compares the influence matrices~$M_h$~\eqref{eq:matrix_M} of the QoI~\eqref{eq:qoi_projectile} for the prediction scenario~$x_\text{pred}$, the tentative validation scenario~$x_\text{val,tent}$, and the aforementioned optimal validation scenario~$x_\text{val}$. 

We observe that the first eigenvector $v_1$, associated with the largest eigenvalue $\lambda_1$, has significantly changed between~$x_\text{val,tent}$ and~$x_\text{val}$ in order to resemble more closely the one of the prediction scenario (especially for the~$u_0$,~$v_0$, and~$g$ components). The second eigenvector also changed notably between the tentative and optimal validation scenarios. Compared to the prediction scenario, the second and third eigenvectors seem to have switched, but this result can be explained by the fact that the second and third eigenvalues of the optimal validation scenario are almost identical. Apart from this, we actually observe a good agreement between the eigenvectors. In other words, the influence matrix of the QoI at~$x_\text{val}$ better mimics the influence matrix of the QoI at~$x_\text{pred}$ than the tentative validation scenario. 

\subsubsection{Optimal Sensors and Observation Functional}
\label{ssec:optimal_sensor_projectile}

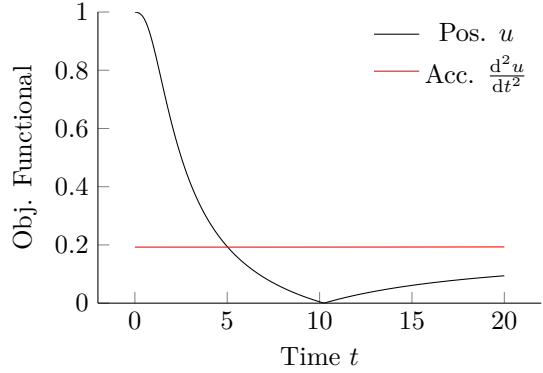
\begin{figure}[t]
    \centering
    \setlength\figureheight{0.25\linewidth}
    \setlength\figurewidth{0.4\linewidth}
    \tikzset{external/export next=false}
    \begin{tikzpicture}[baseline,trim axis left, trim axis right]
\begin{axis}[
    width=0.951\figurewidth,
    height=\figureheight,
    at={(0\figurewidth,0\figureheight)},
    scale only axis,
    xlabel=Time $t$,
    ymin=0,
    ymax=1,
    yminorgrids=true,
    ylabel style={align=center,font=\normalsize},
    ylabel={Obj. Functional},
    axis x line*=bottom,
    axis y line*=left,
    legend style={at={(1,1)}, anchor=north east, draw=none,fill=none},
    ]
    \pgfplotstableread{./Results/projectile/obj_sensor_val.txt}\loadedtable
    
    \addplot [black] table [x=time, y=pos] {\loadedtable};
    \addlegendentry{Pos. $u$}
    
    \addplot [red] table [x=time, y=acc] {\loadedtable};
    \addlegendentry{Acc. $\frac{\diff^2 u}{\diff t^2}$}

\end{axis}
\end{tikzpicture} 
    \caption{Objective functional of Problem~\eqref{eq:obj_function_sensor} associated with the observation functionals~$h_\text{obs}(x_\text{val},\Theta,z)\coloneqq u$ and $h_\text{obs}(x_\text{val},\Theta,z)\coloneqq {\diff^2 u}/{\diff t^2}$.}
    \label{fig:obj_sensor_false_positive}
\end{figure}

We now search for the best observation functional~$h_\text{obs}$ between the functionals~$h_\text{obs}(x_\text{val},\Theta,z) \coloneqq u$ and $h_\text{obs}(x_\text{val},\Theta,z) \coloneqq {\diff^2 u}/{\diff t^2}$, as funtions of time $t$, and the best sensor parameters~$z \coloneqq t$ that should be employed in the validation experiment. In order to do so, we evaluate the objective functional of Problem~\eqref{eq:obj_function_sensor} for both observation functionals and for various times~$t$ under the validation scenario~$x_\text{val}$ computed previously. The results are shown in Figure~\ref{fig:obj_sensor_false_positive}.

As expected, there is a specific time~$t$ for which the influence matrix of~$h_\text{obs}(x_\text{val},\Theta,z) \coloneqq u(t)$ is exactly the same as the influence matrix of~$h_\text{qoi}(x_\text{val},\Theta)$. This particular time~$t$ corresponds to the instant at which the projectile has reached the maximal altitude (as shown in Figure~\ref{fig:comp_scenario}). It is interesting to observe in Figure~\ref{fig:obj_sensor_false_positive} that the objective functionals for the position $u$ and acceleration ${\diff^2 u}/{\diff t^2}$ evolve quite differently in time. On one hand, the influence matrix for the acceleration remains quite different from the influence matrix of the QoI at the validation scenario for all times~$t$. On the other hand, the objective functional~\eqref{eq:obj_function_sensor} for the position~$u$ at small times~$t$ is quite large. This is explained by the fact that the position right after the launch is mainly influenced by the initial position~$u_0$ and initial velocity~$v_0$, which is not the case for the QoI at the validation scenario, see rightmost plot on bottom row in Figure~\ref{fig:comparison_si_qoi_false_positive}. However, a rapid decrease in the objective functional of Problem~\eqref{eq:obj_function_sensor} for the position indicates that the influence of the initial position and velocity (alongside the other parameters) quickly tend to that of the QoI.

For the sake of presenting the whole validation process, and to illustrate the possibility of obtaining a false validation result, we need some experimental observations. We shall use manufactured or synthetic data provided by what we will refer here to as a \emph{fine model} that serves as a surrogate to the physical system of interest. Indeed, the precise source of experimental observations is not directly relevant here, since they do not play any role in the design of the optimal validation experiment. 

\subsubsection{Manufactured Data}
\label{sssec:synthetic_model}

The fine model is provided by the following non-linear ODE~\cite[Section 7.6]{white_fluid_2009} and initial conditions
\begin{subequations}
\label{eq:synthetic_model_projectile}
    \begin{align}
        m\frac{\diff^2 \tilde{u}}{\diff t^2}(t) + \frac{\rho \pi \ell^2 c_D}{8}  \frac{\diff \tilde{u}}{\diff t}(t)\abs*{\frac{\diff \tilde{u}}{\diff t}(t)} = -mg, &\qquad \text{for } t\in(0,T),\\
        \tilde{u}(0) = u_0, \\
        \frac{\diff \tilde{u}}{\diff t}(0) = v_0.
    \end{align}
\end{subequations}
Following~\cite{yang_general_2015}, the definition of the friction coefficient~$c_D$ depends on the Reynolds number~$\mathrm{Re}$~\eqref{eq:Reynoldsnumber} and is defined for laminar regimes, i.e.\ $\mathrm{Re} < 2\times 10^5$, as:
\begin{align}
\label{eq:friction_coefficient}
    c_D = \frac{24}{\mathrm{Re}}\left(1 + 0.15\times\mathrm{Re}^{0.681}\right) + \frac{0.407}{1 + 8710\times\mathrm{Re}^{-1}} .
\end{align}
In case of very low Reynolds numbers ($\mathrm{Re}\ll 1$), the coefficient is well approximated by~$c_D = 24/\mathrm{Re}$. With this approximation, one actually recovers the linear model~\eqref{eq:model_projectile}.

The fine model allows us to obtain measurements of the position~$\tilde{u}$ at any time~$t$. Realizations~$y_\text{exp}(x,z)$ of these pseudo-experimental observations~$Y_\text{exp}(x,z)$ are obtained by sampling
\begin{align}
\label{eq:synthetic_experimental_observations}
    Y_\text{exp}(x,z) &\sim \tilde{u}(x,z)\, \mathrm{N}(1,0.05^2). 
\end{align}
This choice for the noise is arbitrary and corresponds to a multiplicative Gaussian noise with standard deviation~$5\%$. We suppose here that the discretization errors are precisely controlled while solving the ODE~\eqref{eq:synthetic_model_projectile}. Finally, for this fine model, we fix the values of the gravitational constant, the air viscosity, and the air density to~$g=9.81$, $\exp(\mu)=1.8 \times 10^{-5}$, and  $\rho=1.2$, respectively. 

\subsubsection{Validation Analysis}
\label{sssec:validation_analysis}

\begin{figure}[t]
    \centering
    \setlength\figureheight{0.25\linewidth}
    \setlength\figurewidth{0.25\linewidth}
    \setlength{\tabcolsep}{15pt}
    \begin{tabular}{lcr}
        \tikzset{external/export next=false}
        \begin{tikzpicture}[baseline,trim axis left, trim axis right]
\begin{axis}[
    width=0.951\figurewidth,
    height=\figureheight,
    at={(0\figurewidth,0\figureheight)},
    scale only axis,
    xlabel=Time $t$,
    ylabel = Position $u$,
    title = Prediction Scenario,
    yminorgrids=true,
    axis x line*=bottom,
    axis y line*=left,
    ]
    \pgfplotstableread{./Results/projectile/pos_pred.txt}\datamodel
    \pgfplotstableread{./Results/projectile/pos_pred_exact.txt} \dataexact
    
    \addplot [blue, dashed, name path=MeanPos] table [x=time, y=mean_pos] {\datamodel};

    \addplot [draw=none, name path=LowerPos,forget plot] table [x=time, y=lower_pos] {\datamodel};

    \addplot [draw=none, name path=UpperPos,forget plot] table [x=time, y=upper_pos] {\datamodel};

    \addplot [blue,fill opacity=0.3,forget plot] fill between [of=LowerPos and UpperPos];

    \addplot [red, name path=MeanPosExact] table [x=time, y=mean_pos] {\dataexact};

    \addplot [draw=none, name path=LowerPosExact,forget plot] table [x=time, y=lower_pos] {\dataexact};

    \addplot [draw=none, name path=UpperPosExact,forget plot] table [x=time, y=upper_pos] {\dataexact};

    \addplot [red,fill opacity=0.3,forget plot] fill between [of=LowerPosExact and UpperPosExact];
    
    \draw [black,loosely dashed] (axis cs:10.15,\pgfkeysvalueof{/pgfplots/ymin}) -- (axis cs:10.15,\pgfkeysvalueof{/pgfplots/ymax});

\end{axis}
\end{tikzpicture} &
        \tikzset{external/export next=false}
        \begin{tikzpicture}[baseline,trim axis left, trim axis right]
\begin{axis}[
    width=0.951\figurewidth,
    height=\figureheight,
    at={(0\figurewidth,0\figureheight)},
    scale only axis,
    xlabel=Time $t$,
    title = Tentative Validation Scenario,
    yminorgrids=true,
    axis x line*=bottom,
    axis y line*=left,
    ]
    \pgfplotstableread{./Results/projectile/pos_init.txt}\datamodel
    \pgfplotstableread{./Results/projectile/pos_init_exact.txt} \dataexact
    
    \addplot [blue, dashed, name path=MeanPos] table [x=time, y=mean_pos] {\datamodel};

    \addplot [draw=none, name path=LowerPos,forget plot] table [x=time, y=lower_pos] {\datamodel};

    \addplot [draw=none, name path=UpperPos,forget plot] table [x=time, y=upper_pos] {\datamodel};

    \addplot [blue,fill opacity=0.3,forget plot] fill between [of=LowerPos and UpperPos];

    \addplot [red, name path=MeanPosExact] table [x=time, y=mean_pos] {\dataexact};

    \addplot [draw=none, name path=LowerPosExact,forget plot] table [x=time, y=lower_pos] {\dataexact};

    \addplot [draw=none, name path=UpperPosExact,forget plot] table [x=time, y=upper_pos] {\dataexact};

    \addplot [red,fill opacity=0.3,forget plot] fill between [of=LowerPosExact and UpperPosExact];

    \draw [black,loosely dashed] (axis cs:2.0301,\pgfkeysvalueof{/pgfplots/ymin}) -- (axis cs:2.0301,\pgfkeysvalueof{/pgfplots/ymax});

\end{axis}
\end{tikzpicture} &
        \tikzset{external/export next=false}
        \begin{tikzpicture}[baseline,trim axis left, trim axis right]
\begin{axis}[
    width=0.951\figurewidth,
    height=\figureheight,
    at={(0\figurewidth,0\figureheight)},
    scale only axis,
    xlabel=Time $t$,
    title = Validation Scenario,
    yminorgrids=true,
    axis x line*=bottom,
    axis y line*=left,
    ]
    \pgfplotstableread{./Results/projectile/pos_val.txt}\datamodel
    \pgfplotstableread{./Results/projectile/pos_val_exact.txt} \dataexact
    
    \addplot [blue, dashed, name path=MeanPos] table [x=time, y=mean_pos] {\datamodel};

    \addplot [draw=none, name path=LowerPos,forget plot] table [x=time, y=lower_pos] {\datamodel};

    \addplot [draw=none, name path=UpperPos,forget plot] table [x=time, y=upper_pos] {\datamodel};

    \addplot [blue,fill opacity=0.3,forget plot] fill between [of=LowerPos and UpperPos];

    \addplot [red, name path=MeanPosExact] table [x=time, y=mean_pos] {\dataexact};

    \addplot [draw=none, name path=LowerPosExact,forget plot] table [x=time, y=lower_pos] {\dataexact};

    \addplot [draw=none, name path=UpperPosExact,forget plot] table [x=time, y=upper_pos] {\dataexact};

    \addplot [red,fill opacity=0.3,forget plot] fill between [of=LowerPosExact and UpperPosExact];

    \draw [black,loosely dashed] (axis cs:10.25,\pgfkeysvalueof{/pgfplots/ymin}) -- (axis cs:10.25,\pgfkeysvalueof{/pgfplots/ymax});

\end{axis}
\end{tikzpicture} \\
        \tikzset{external/export next=false}
        \begin{tikzpicture}[baseline,trim axis left, trim axis right]
\begin{axis}[
    width=0.951\figurewidth,
    height=\figureheight,
    at={(0\figurewidth,0\figureheight)},
    scale only axis,
    xlabel= Observation Discrepancy~$\abs*{e}$,
    ymin=0,
    ymax=1,
    yminorgrids=true,
    ylabel=$F_{\abs*{E}}$,
    axis x line*=bottom,
    axis y line*=left,
    legend style={at={(1,0.5)}, anchor=west, draw=none,fill=none},
    ]
    \pgfplotstableread{./Results/projectile/metric_pred.txt}\loadedtable
    
    \addplot [black] table [x=discrepancy, y=cdf] {\loadedtable};

\end{axis}
\end{tikzpicture} &
        \tikzset{external/export next=false}
        \begin{tikzpicture}[baseline,trim axis left, trim axis right]
\begin{axis}[
    width=0.951\figurewidth,
    height=\figureheight,
    at={(0\figurewidth,0\figureheight)},
    scale only axis,
    xlabel= Observation Discrepancy~$\abs*{e}$,
    ymin=0,
    ymax=1,
    yminorgrids=true,
    axis x line*=bottom,
    axis y line*=left,
    legend style={at={(1,0.5)}, anchor=west, draw=none,fill=none},
    ]
    \pgfplotstableread{./Results/projectile/metric_init.txt}\loadedtable
    
    \addplot [black] table [x=discrepancy, y=cdf] {\loadedtable};

\end{axis}
\end{tikzpicture} &
        \tikzset{external/export next=false}
        \begin{tikzpicture}[baseline,trim axis left, trim axis right]
\begin{axis}[
    width=0.951\figurewidth,
    height=\figureheight,
    at={(0\figurewidth,0\figureheight)},
    scale only axis,
    xlabel= Observation Discrepancy~$\abs*{e}$,
    ymin=0,
    ymax=1,
    yminorgrids=true,
    axis x line*=bottom,
    axis y line*=left,
    legend style={at={(1,0.5)}, anchor=west, draw=none,fill=none},
    ]
    \pgfplotstableread{./Results/projectile/metric_val.txt}\loadedtable
    
    \addplot [black] table [x=discrepancy, y=cdf] {\loadedtable};

\end{axis}
\end{tikzpicture} 
    \end{tabular}
    \caption{Position~$u$ (top row) and discrepancies (bottom row) for the prediction scenario (left column), tentative validation scenario (center column), and the optimal validation scenario (right column). The experimental observations are in red and the model observations are in blue. The shaded areas represent the 95\% confidence interval for both the experimental and model observations. The absolute value of the discrepancies~\eqref{eq:discrepancy_model} are computed at the sensor location indicated by the black dashed lines.}
    \label{fig:comp_scenario}
\end{figure}
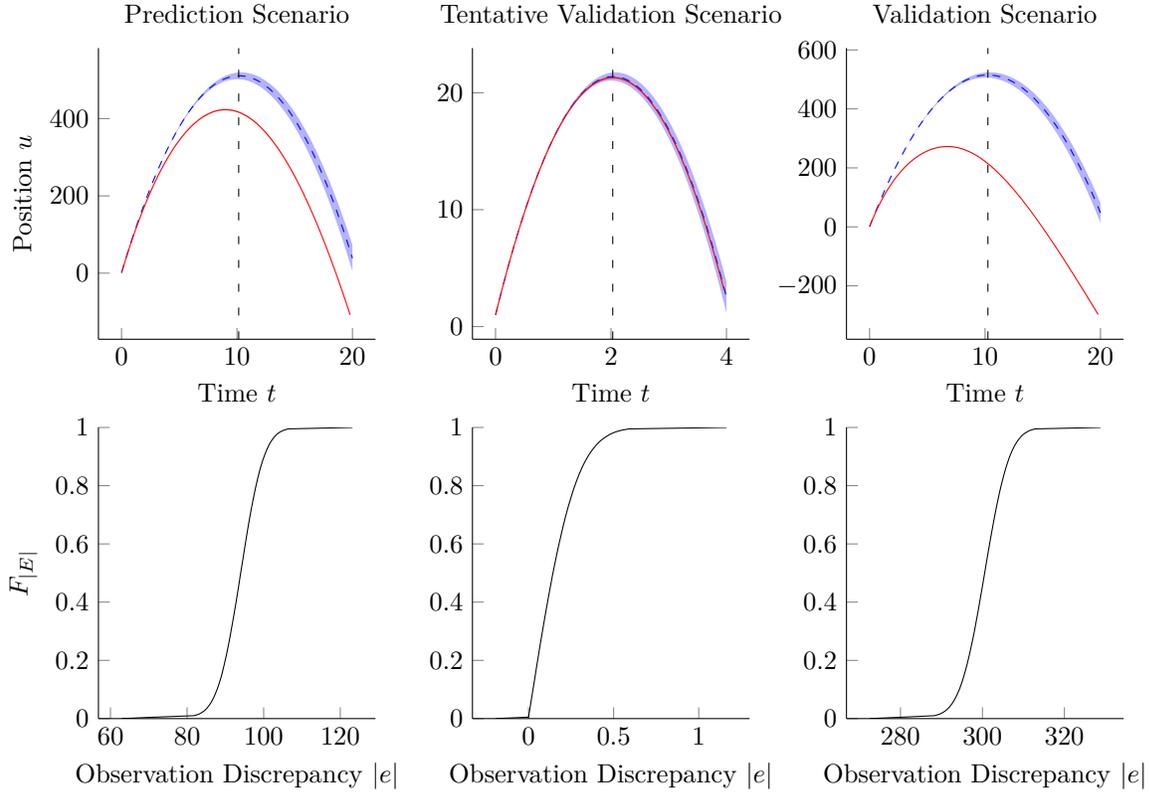

We now compare the trajectories obtained with  Model~\eqref{eq:model_projectile} and Model~\eqref{eq:synthetic_model_projectile} for the prediction scenario~$x_\text{pred}$, the tentative validation scenario~$x_\text{val,tent}$, and the optimal validation scenario~$x_\text{val}$. Moreover, we compute the discrepancy~\eqref{eq:discrepancy_model} for the three scenarios at their respective optimal sensor parameter~$z_\text{val}$, which corresponds to the time the projectile reaches the maximal altitude. The trajectories and distributions~$F_{\abs*{E}}$ of the discrepancy~$E$ are shown in Figure~\ref{fig:comp_scenario}. It is important to mention that the distribution of the discrepancy~$E$ and the trajectory of the physical system (red curve) cannot in general be accessed for the prediction scenario since experimental data are not available in that case. We can produce those plots here only because the measurements are obtained with the help of the model~\eqref{eq:synthetic_model_projectile}.

We observe that our model~\eqref{eq:model_projectile} poorly captures the maximum altitude reached by the projectile at the prediction scenario, since there is a significant difference between the red and blue curves, as shown in the top left plot of Figure~\ref{fig:comp_scenario}. The friction forces in the model are in fact underestimated. This mismatch between the model and the system of interest is reflected in the distribution~$F_{\abs*{E}}$ of the discrepancy: the probability that the absolute error on the maximal position be less than 80 is close to zero. In other words, one is almost sure that the absolute error on the maximal position will be between 80 and 110. The model for the prediction of the maximal altitude under the prediction scenario~$x_\text{pred}$ can thus be considered invalid. Although this information is not available in practice, the validation process should inform us beforehand on the accuracy of the model predictions.

Under the tentative validation scenario~$x_\text{val,tent}$, we observe a good agreement between the position of the projectile for the model and for the system of interest. This can be explained by the fact that the friction force is in this case negligible with respect to the inertial and gravitational forces. Since the only difference between the model and the system of interest is due to the modeling of the friction force, the discrepancy~$E$, computed at the optimal sensor point~$z=t\approx 2$, remains small. Its distribution~$F_{\abs*{E}}$ indicates that the observational error is almost surely less than 0.5. Depending on the tolerance~$\varepsilon$ and threshold~$\eta$ on the validation metric, the model would likely have been deemed valid. Hence, the model is not appropriate for the prediction of the QoI at the prediction scenario and the validation process using this tentative validation scenario would have given a false positive.

In the case of the optimal validation scenario~$x_\text{val}$, we observe a significant departure between the model and the system of interest. The optimal validation scenario seems to capture the fact that the friction forces in our model are underestimated. With the optimal sensor point previously computed, the discrepancy indicates that the error on the position is almost surely between 280 and 310. The model would likely have failed the validation test, which better reflects the real predictive capabilities of the model under the prediction scenario.  

We would like to emphasize here that the proposed methodology to design an optimal validation experiment does not guarantee that the outcome of the validation process will always provide one with a correct answer. It only ascertains that the influence matrix of the observation at the validation scenario is as much as possible comparable to the influence matrix of the QoI at the prediction scenario. In this example, it happens that the conditions of the optimal validation experiment allow the model for capturing the relative importance of the friction forces with respect to the other forces, as in the prediction scenario. In other words, since the friction forces are not correctly modeled in the validation setting, one may conclude that it could also be the case for the prediction scenario, which is indeed true here.

\subsection{Contaminant Transport Problem}
\label{ssec:contaminant_problem}

The second example consists in a problem of pollutant transport in a fictitious river where we wish to analyze the impact of the design of a new factory upstream. We consider two QoIs, both representing the mean concentration of contaminant in a specific region of the domain. The model we wish to validate is a 2D linear steady-state diffusion-advection equation that governs the concentration~$\conc=\conc(z)$ of the contaminant in the river:
\begin{subequations} 
\label{eq:model_contaminant}
\begin{align}
    -\nabla\cdot(\exp(k)\nabla \conc) + \nabla \cdot(\vel\conc) =0,  &\qquad \text{in}\ \Omega,\\
    \conc = \conc_D, &\qquad \text{on}\ \Gamma_\text{west},\\
    -n \cdot \exp(k)\nabla\conc = 0, &\qquad \text{on}\ \partial \Omega \setminus \Gamma_\text{west},
\end{align}
\end{subequations}
where $n$ denotes the outward unit normal to the domain boundary, $\exp(k)$ is the diffusivity coefficient, and $\vel$ is the advection velocity. For the sake of simplicity, we consider the velocity field~$\vel$ to be given and known, so that it is not considered a parameter in this study (see~\ref{ann:velocity}). 

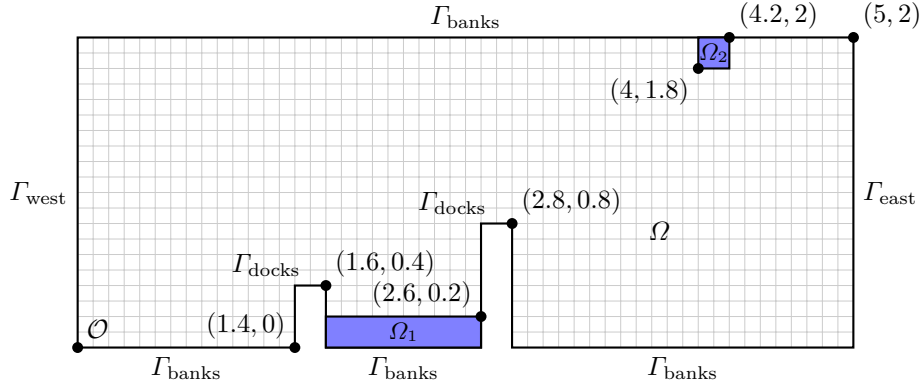
\begin{figure}[t]
    \centering
    \def\figurescale{0.75} 
    \tikzset{external/export next=false}
    \pgfmathsetlengthmacro{\nodedistance}{2cm*\figurescale}
\pgfmathsetlengthmacro{\twidth}{4cm*\figurescale}
\begin{tikzpicture}[scale=\figurescale,node distance=\nodedistance,baseline]

\begin{scope}[xscale=2.75,yscale=2.75]

\node [coordinate, label=above right:{$\mathcal{O}$}] (origin) at (0,0)  {};
\node [coordinate, label=above left:{$(1.4,0)$}] (v2) at (1.4,0)  {};
\node [coordinate, label=above right:{$(1.6,0.4)$}] (v4) at (1.6,0.4)  {};

\node [coordinate] (v6) at (2.6,0)  {};
\node [coordinate, label=above right:{$(2.8,0.8)$}] (v8) at (2.8,0.8)  {};
\node [coordinate, label=above right:{$(5,2)$}] (v11) at (5,2)  {};

\draw [thick] (origin) -- (v2) node[pos=0.5,below]  {$\Gamma_\text{banks}$} -- 
(1.4,0.4) -- node[pos=0.5,above left] {$\Gamma_\text{docks}$}
(v4) -- 
(1.6,0) -- 
(v6) node[pos=0.5,below] {$\Gamma_\text{banks}$} -- 
(2.6,0.8) -- node[pos=0.5,above left] {$\Gamma_\text{docks}$}
(v8) -- 
(2.8,0) -- 
(5,0) node[pos=0.5,below] {$\Gamma_\text{banks}$} -- 
(v11)node[pos=0.5,right] {$\Gamma_\text{east}$} -- 
(0,2) node[pos=0.5,above] {$\Gamma_\text{banks}$} -- 
cycle node[pos=0.5,left] {$\Gamma_\text{west}$} ;

\node [coordinate, label=above left:{$(2.6,0.2)$}] (vqoi1_2) at (2.6,0.2) {};

\draw [thick,fill=blue!50] (1.6,0) rectangle (2.6,0.2) node[pos=.5] {\small$\Omega_1$};

\node [coordinate, label=below left:{$(4,1.8)$}] (vqoi2_1) at (4,1.8) {};
\node [coordinate, label=above right:{$(4.2,2)$}] (vqoi2_2) at (4.2,2) {};

\draw [thick,fill=blue!50] (vqoi2_1) rectangle (vqoi2_2) node[pos=.5] {\small$\Omega_2$};

\fill[black,inner sep=2pt] (origin) circle (1pt);
\fill[black,inner sep=2pt] (v2) circle (1pt);
\fill[black,inner sep=2pt] (v4) circle (1pt);
\fill[black,inner sep=2pt] (v8) circle (1pt);
\fill[black,inner sep=2pt] (v11) circle (1pt);
\fill[black,inner sep=2pt] (vqoi1_2) circle (1pt);
\fill[black,inner sep=2pt] (vqoi2_1) circle (1pt);
\fill[black,inner sep=2pt] (vqoi2_2) circle (1pt);

\node (domain) at (3.75,0.75) {$\Omega$};

\begin{scope}[on background layer]
\clip (origin) -- (v2)  -- 
(1.4,0.4) -- 
(v4) -- 
(1.6,0) -- 
(v6) -- 
(2.6,0.8) -- 
(v8) -- 
(2.8,0) -- 
(5,0) -- 
(v11) -- 
(0,2) -- 
cycle;
\draw[step=0.1,line width = 0.1pt, color=black!20] (0,0) grid (5,2);
\end{scope}

\end{scope}

\end{tikzpicture}     
    \caption{Domain~$\Omega$ and its boundaries for the contaminant transport model. The regions for which we seek the mean pollutant concentration are indicated by~$\Omega_1$ and~$\Omega_2$ (respectively employed for the definition of the QoIs). The grid is used to indicate the regions~$\Omega_\text{obs}$ in which the mean concentration of pollutant can be observed.}
    \label{fig:domain}
\end{figure}

The control parameters~$x$ consist in the parameters that characterize the Dirichlet condition~$\conc_D$, to be described later, and the model parameter is only $\theta = k$. We suppose here that the diffusivity parameter~$k$ is well-known so that~$\Theta \sim \delta(k_0)$, with~$\delta$ the Dirac measure and~$k_0=-2\ln(10)$. The sensor parameters are the Cartesian coordinates~$z=(z_1,z_2)$. A sketch of the geometry is provided in Figure~\ref{fig:domain}. The domain consists in an inlet and an outlet on the west and east boundaries, respectively. Two docks hinder the flow around the first region of interest, denoted by~$\Omega_1$. The second region of interest $\Omega_2$ is shown in Figure~\ref{fig:domain}. We note that the particular dimensions of the domain~$\Omega$ and the particular value of the model parameter~$k$ are arbitrarily chosen, as the example is for illustrative purpose only.

To validate the model, we imagine one can perform an experiment consisting in injecting a small and controlled amount of contaminant upstream. The Dirichlet condition~$\conc_D$ is therefore used to describe both the pollutant release from the factory and the validation experiment. We parametrize~$\conc_D$ in terms of the mollifier centered at~$z_2=z_0$, of length~$L$ and  intensity~$c$, as follows:
\begin{align}
\label{eq:mollifier}
\conc_D(z_0,L,c,z_2) = 
\begin{cases} 
    c \exp\Big(\big(\abs*{\frac{z_2 - z_0}{L}}-1\big)^{-1}\Big), & 
    \text{if $\abs*{\frac{z_2 - z_0}{L}} < 1$},\\
    0, & \text{otherwise}.
\end{cases}
\end{align}
The control parameters in this application are given by~$x=(z_0,L,c)$. This parametrization is sufficiently rich to describe both the pollutant release from the factory and a wide range of validation scenarios, while involving a limited number of parameters. Moreover, the use of above mollifier is attractive thanks to its regularity properties, since we need to compute the gradient of functionals with respect to the control parameter~$x$ (as explained in~\ref{ann:gradient}). 

We can now specify both the prediction scenario~$x_\text{pred}$ and the controlled environment~$\mathcal{X}_\text{lab}$. The prediction scenario consists in a description of the possible pollutant profile from the factory to be built upstream
and is provided in Table~\ref{tab:prediction_scenario_pollutant}. The controlled environment~$\mathcal{X}_\text{lab}$ includes all validation scenarios that can actually be performed. We suppose here that the pollutant can be injected in a controlled manner anywhere along the boundary~$\Gamma_\text{west}$, while the length~$L$ and intensity~$c$ can only take limited values. The intervals for the three control parameters~$z_0$, $L$, and~$c$ are also provided in Table~\ref{tab:prediction_scenario_pollutant}. 

\begin{table}[t]
\centering
\caption{Description of the prediction scenario~$x_\text{pred}$ and of the controlled environment~$\mathcal{X}_\text{lab}$ for the pollutant transport model.}
\label{tab:prediction_scenario_pollutant}
\vspace{2pt}
\begin{tabular}{@{}cccc@{}}
\toprule
 &
  \textbf{Position}~$z_0$ &
  \textbf{Length}~$L$ &
  \textbf{Intensity}~$c$\\ \midrule
\begin{tabular}[c]{@{}c@{}}
\textbf{Prediction}\\ 
\textbf{Scenario}~$x_\text{pred}$
\end{tabular} & 0.75 & 0.6 & 2
\\  \midrule
\begin{tabular}[c]{@{}c@{}}
\textbf{Controlled} \\ 
\textbf{Environment}~$\mathcal{X}_\text{lab}$
\end{tabular} & [0.1,1.9] & [0.05,0.1] & [0.5,1] \\\bottomrule
\end{tabular}
\end{table}

We also consider the two QoIs given by
\begin{align*}
\label{eq:qoi_pollutant}
    h_{\text{qoi},i}(x,\theta) = \frac{1}{\abs*{\Omega_i}} \int_{\Omega_i} \conc(x,\theta,z)\, \diff z, \qquad \text{for}\ i=1,2,  
\end{align*}
where~$\Omega_i$, $i=1,2$, are the two sub-regions shown in Figure~\ref{fig:domain}. 

\subsubsection{Optimal Validation Scenarios}

Now that the prediction scenario~$x_\text{pred}$, quantities of interest~$h_{\text{qoi},i}$, and controlled environment~$\mathcal{X}_\text{lab}$ are prescribed, we can compute the optimal validation scenarios with respect to the two QoIs. We thus solve Problem~\eqref{eq:obj_function_control} for each QoI. The concentration of the pollutant for the prediction scenario~$x_\text{pred}$ and the optimal validation scenarios with respect to the first and second QoI are all shown in Figure~\ref{fig:pollutant_scenario}.

\newcommand{\colorbarcustom}[2]{\begin{tikzpicture}[baseline, trim axis left, trim axis right]
\begin{axis}[
    hide axis,
    scale only axis,
    height=0pt,
    width=0pt,
    colormap/viridis,
    colorbar,
    point meta min=#1,
    point meta max=#2,
    colorbar style={
        at={(0,0)},
        anchor=south west,
        height=\figureheight,
    }
]
\addplot [draw=none] coordinates {(0,0) (1,1)};
\end{axis}
\end{tikzpicture}
}

\begin{figure}[t]
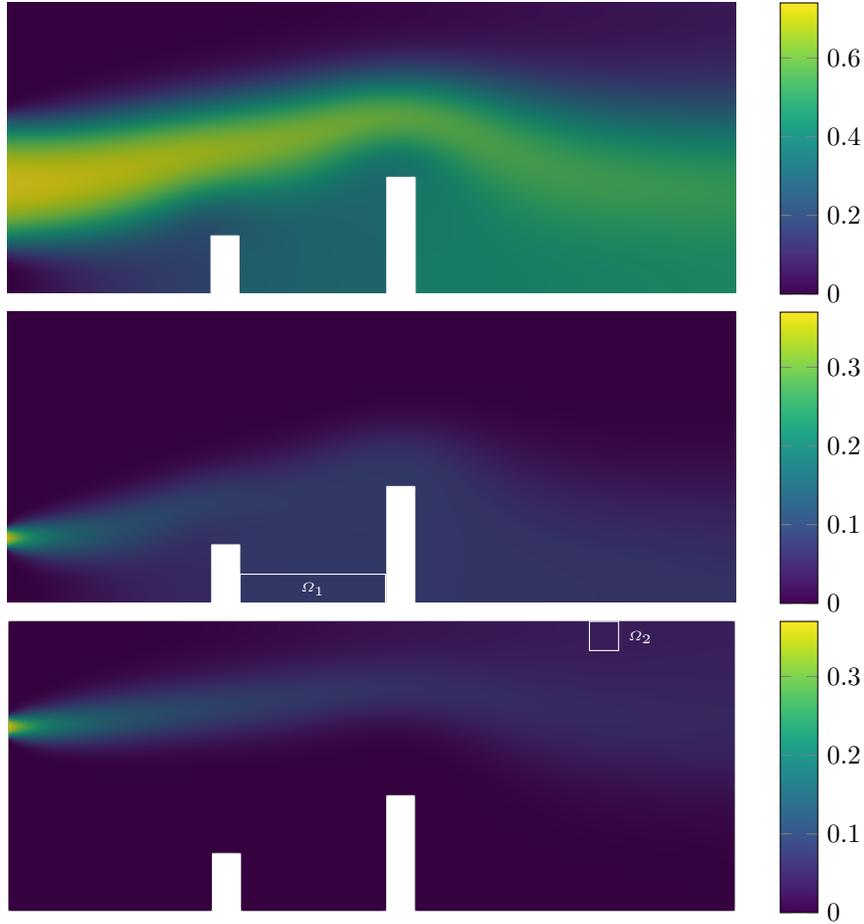

    \centering
    \setlength\figureheight{0.25\linewidth}
    \setlength\figurewidth{0.05\linewidth}
    \setlength{\tabcolsep}{5pt}
    \begin{tabular}{rl}
        \tikzset{external/export next=false}
        \input{./Figures/pollutant/pred_scenario.tikz}
        &
        \tikzset{external/export next=false}
        \colorbarcustom{0}{0.74}  \\
        \tikzset{external/export next=false}
        \input{./Figures/pollutant/optimal_val_scenario_qoi1.tikz}
        &
        \tikzset{external/export next=false}
        \colorbarcustom{0}{0.37} \\
        \tikzset{external/export next=false}
        \input{./Figures/pollutant/optimal_val_scenario_qoi2.tikz}
        &
        \tikzset{external/export next=false}
        \colorbarcustom{0}{0.37}
    \end{tabular}
    \caption{Concentration of the pollutant for the prediction and optimal validation scenarios: (top) prediction scenario, (center) optimal validation scenario associated with~$h_{\text{qoi},1}$, (bottom) optimal validation scenario associated with~$h_{\text{qoi},2}$.}
    \label{fig:pollutant_scenario}
\end{figure}

We observe that the optimal validation scenarios are highly sensitive to the particular QoI we wish to predict since the location~$z_0$ of the pollutant varies considerably between the two QoIs. This result is important as it highlights the fact that it is the combination of the prediction scenario and the QoI one wants to predict that should dictate the validation scenario. From the prediction scenario, we observe that the pollutant enters the domain~$\Omega_1$ associated with the first QoI, while very little pollutant reaches the second domain~$\Omega_2$ for the second QoI. We reiterate here that the objective is not to find the validation scenario~$x_\text{val}$ such that~$h_\text{qoi}(x_\text{val},\Theta) \approx h_\text{qoi}(x_\text{pred},\Theta)$, but the scenario such that~$M_{h_\text{qoi}}(x_\text{val}) \approx M_{h_\text{qoi}}(x_\text{pred})$. 

\begin{figure}[t]
    \centering
    \setlength\figureheight{0.2\linewidth}
    \setlength\figurewidth{0.2\linewidth}
    \setlength{\tabcolsep}{25pt}
    \begin{tabular}{lr}
        \tikzset{external/export next=false}
        \begin{tikzpicture}[baseline,trim axis left, trim axis right]
\begin{axis}[
    width=0.951\figurewidth,
    height=\figureheight,
    at={(0\figurewidth,0\figureheight)},
    scale only axis,
    xlabel=Index,
    xtick={1,2,3,4,5,6},
    ymin=1e-19,
    ymax=1e4,
    ylabel=$\lambda$,
    ymode = log,
    axis x line*=bottom,
    axis y line*=left,
    legend style={at={(0,1)}, anchor=north west, draw=none,fill=none},
    title={$M_{h_\text{qoi,1}}(x)$},
    ]
    \pgfplotstableread{./Results/pollutant/eigenvalue.txt}\loadedtable
    
    \addplot [mark options={fill=black},only marks] table [x=index,y=lambda_pred_center] {\loadedtable};
    \addplot [red,mark=square*,only marks] table [x=index,y=lambda_val_center] {\loadedtable};
    
\end{axis}
\end{tikzpicture} &
        \tikzset{external/export next=false}
        \begin{tikzpicture}[baseline,trim axis left, trim axis right]
\begin{axis}[
    width=0.951\figurewidth,
    height=\figureheight,
    at={(0\figurewidth,0\figureheight)},
    scale only axis,
    xlabel=Index,
    xtick={1,2,3,4,5,6},
    ymin=1e-19,
    ymax=1e4,
    ylabel=$\lambda$,
    ymode = log,
    axis x line*=bottom,
    axis y line*=left,
    legend style={at={(0,1)}, anchor=north west, draw=none,fill=none},
    title={$M_{h_\text{qoi,2}}(x)$},
    ]
    \pgfplotstableread{./Results/pollutant/eigenvalue.txt}\loadedtable
    
    \addplot [black,only marks] table [x=index,y=lambda_pred_top] {\loadedtable};
    \addplot [red,mark=square*,only marks] table [x=index,y=lambda_val_top] {\loadedtable};
    
\end{axis}
\end{tikzpicture}   \\
        \tikzset{external/export next=false}
        \begin{tikzpicture}[baseline,trim axis left, trim axis right]
\begin{axis}[
    width=0.951\figurewidth,
    height=\figureheight,
    at={(0\figurewidth,0\figureheight)},
    scale only axis,
    xtick={1,2,3,4,5,6},
    xticklabels={$z_0$,$L$,$c$,$k$},
    ymin=-1,
    ymax=1,
    yminorgrids=true,
    axis x line*=bottom,
    axis y line*=left,
    legend style={at={(0,1)}, anchor=north west, draw=none,fill=none},
    ]
    \pgfplotstableread{./Results/pollutant/eigenvector_pred_center.txt}\datapred
    \pgfplotstableread{./Results/pollutant/eigenvector_val_center.txt}\dataval

    \addplot [black,mark=*] table [x=index,y=v1] {\datapred};
    \addplot [red,mark=square*] table [x=index,y=v1] {\dataval};

\end{axis}
\end{tikzpicture} &
        \tikzset{external/export next=false}
        \begin{tikzpicture}[baseline,trim axis left, trim axis right]
\begin{axis}[
    width=0.951\figurewidth,
    height=\figureheight,
    at={(0\figurewidth,0\figureheight)},
    scale only axis,
    xtick={1,2,3,4,5,6},
    xticklabels={$z_0$,$L$,$c$,$k$},
    ymin=-1,
    ymax=1,
    yminorgrids=true,
    axis x line*=bottom,
    axis y line*=left,
    legend style={at={(0,1)}, anchor=north west, draw=none,fill=none},
    ]
    \pgfplotstableread{./Results/pollutant/eigenvector_pred_top.txt}\datapred
    \pgfplotstableread{./Results/pollutant/eigenvector_val_top.txt}\dataval

    \addplot [black,mark=*] table [x=index,y=v1] {\datapred};
    \addplot [red,mark=square*] table [x=index,y=v1] {\dataval};

\end{axis}
\end{tikzpicture}
    \end{tabular}
    \caption{Eigenvalues (top row) and eigenvectors (bottom row) associated with the influence matrices for the prediction scenario (black circles) and for the validation scenario (red squares). The left and right columns display the sensitivity indices related to the first and second QoIs, respectively. For the sake of clarity, only the first (and only relevant) eigenvector is displayed for each influence matrix.}
    \label{fig:si_pollutant}
\end{figure}
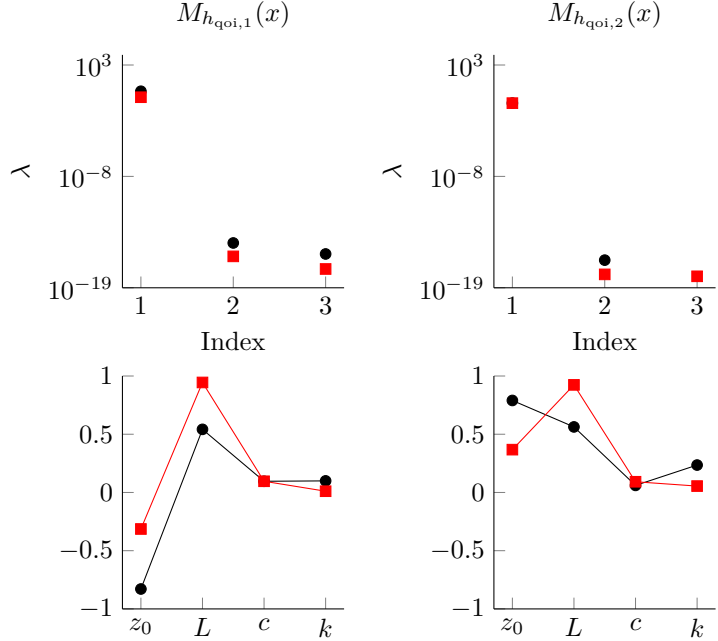

The influence matrices~$M_{h_\text{qoi}}(x_\text{val})$ and $M_{h_\text{qoi}}(x_\text{pred})$ are characterized by their eigenvalues and eigenvectors shown in Figure~\ref{fig:si_pollutant}. The absence of uncertainty in the model parameter~$\theta=k$ implies that the influence matrices~$M_{h_\text{qoi}}$ should be of rank one, since the covariance in Equation~\eqref{eq:matrix_M} is null in this case. This is clearly reflected in the numerical results, where the first eigenvalue of each influence matrix are several orders of magnitude higher than the others.

Concerning the first QoI, we observe that the magnitudes of the eigenvalues are slightly smaller for the validation than those for the prediction. This can be explained by the fact that the overall magnitude of the pollutant is smaller in the validation setting because of the constraint on the controlled environment~$\mathcal{X}_\text{lab}$. We also observe that the eigenvector for the optimal validation scenario tends to match as closely as possible the eigenvector of the prediction scenario. One can explain this difference by remarking that the size $L$ of the region in which the pollutant is injected should have more influence in the validation scenario since the quantity of injected pollutant is smaller than for the prediction scenario. The optimal validation scenario actually succeeds in capturing the fact that the intensity~$c$ of the source of pollutant and the diffusivity~$k$ are not very influential in determining the mean pollutant concentration in~$\Omega_1$. 

Concerning the second QoI, we observe a very good agreement between the eigenvalues of the influence matrix at the validation scenario and those at the prediction scenario. As with the first QoI, the eigenvector for the validation scenario tends to match as closely as possible the eigenvector for the prediction scenario.

\subsubsection{Positioning of the Sensors}

\begin{figure}[t]
    \centering
    \setlength\figureheight{0.25\linewidth}
    \setlength{\tabcolsep}{5pt}
    \begin{tabular}{rl}
        \tikzset{external/export next=false}
        \begin{tikzpicture}[baseline,trim axis left, trim axis right,scale=\figureheight/2cm]

\node [coordinate] (origin) at (0,0)  {};
\node [coordinate] (v2) at (1.4,0)  {};
\node [coordinate] (v4) at (1.6,0.4)  {};

\node [coordinate] (v6) at (2.6,0)  {};
\node [coordinate] (v8) at (2.8,0.8)  {};
\node [coordinate] (v11) at (5,2)  {};

\draw [draw=black, thick] (origin) -- (v2)  -- 
(1.4,0.4) -- 
(v4) -- 
(1.6,0) -- 
(v6) -- 
(2.6,0.8) -- 
(v8) -- 
(2.8,0) -- 
(5,0) -- 
(v11) -- 
(0,2) -- 
cycle;

\begin{scope}[on background layer]
\clip (origin) -- (v2)  -- 
(1.4,0.4) -- 
(v4) -- 
(1.6,0) -- 
(v6) -- 
(2.6,0.8) -- 
(v8) -- 
(2.8,0) -- 
(5,0) -- 
(v11) -- 
(0,2) -- 
cycle;
\begin{axis}[
    width=5cm,
    height=2cm,
    at={(0\figurewidth,0\figureheight)},
    scale only axis,
    view={0}{90},
    xmin=0,
    xmax=5,
    ymin=0,
    ymax=2,
    axis equal image,
    hide axis,
    colormap/viridis,
    point meta min=0,
    point meta max=1,
    ]
    \addplot3 [surf,mesh/rows=21,shader=flat corner,draw=black, ultra thin] table[col sep=comma,x=x, y=y, z=obj] {./Results/pollutant/obj_obs_center_new_mesh.csv};
\end{axis}  
\end{scope}

\draw [white] (1.6,0) rectangle (2.6,0.2) node[pos=.5] {\tiny $\Omega_1$};

\end{tikzpicture} &
        \tikzset{external/export next=false}
        \colorbarcustom{0}{1} \\
        \tikzset{external/export next=false}
        \begin{tikzpicture}[baseline,trim axis left, trim axis right,scale=\figureheight/2cm]

\node [coordinate] (origin) at (0,0)  {};
\node [coordinate] (v2) at (1.4,0)  {};
\node [coordinate] (v4) at (1.6,0.4)  {};

\node [coordinate] (v6) at (2.6,0)  {};
\node [coordinate] (v8) at (2.8,0.8)  {};
\node [coordinate] (v11) at (5,2)  {};

\draw [draw=black, thick] (origin) -- (v2)  -- 
(1.4,0.4) -- 
(v4) -- 
(1.6,0) -- 
(v6) -- 
(2.6,0.8) -- 
(v8) -- 
(2.8,0) -- 
(5,0) -- 
(v11) -- 
(0,2) -- 
cycle;

\begin{scope}[on background layer]
\clip (origin) -- (v2)  -- 
(1.4,0.4) -- 
(v4) -- 
(1.6,0) -- 
(v6) -- 
(2.6,0.8) -- 
(v8) -- 
(2.8,0) -- 
(5,0) -- 
(v11) -- 
(0,2) -- 
cycle;
\begin{axis}[
    width=5cm,
    height=2cm,
    at={(0\figurewidth,0\figureheight)},
    scale only axis,
    view={0}{90},
    xmin=0,
    xmax=5,
    ymin=0,
    ymax=2,
    axis equal image,
    hide axis,
    colormap/viridis,
    point meta min=0,
    point meta max=1,
    ]
    \addplot3 [surf,mesh/rows=21,shader=flat corner,draw=black, ultra thin] table[col sep=comma,x=x, y=y, z=obj] {./Results/pollutant/obj_obs_top_new2.csv};
\end{axis}  
\end{scope}

\draw [white] (4,1.8) rectangle (4.2,2) node[below right] {\tiny $\Omega_2$};

\end{tikzpicture} &
        \tikzset{external/export next=false}
        \colorbarcustom{0}{1}
    \end{tabular}
    \caption{Value of the objective function in~\eqref{eq:obj_function_sensor} for the first QoI (top) and second QoI (bottom).}
    \label{fig:sensor_pollutant}
\end{figure}
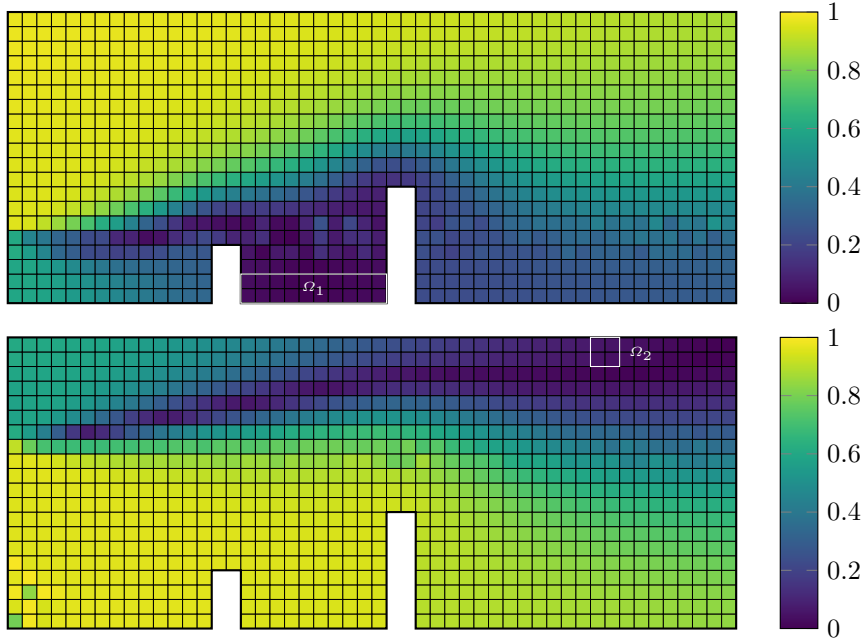

Once the optimal validation scenario has been computed for each quantity of interest considered, we tackle the last step of our methodology, namely the optimal positioning of a sensor. For the pollutant transport problem, we consider only one type of observation functional~$h_\text{obs}$
\begin{align}
    h_\text{obs}(x,\theta,z) = \frac{1}{\abs*{\Omega_\text{obs}(z)}} \int_{\Omega_\text{obs}(z)} \phi(x,\theta,w) \, \diff w,
\end{align}
where~$\Omega_\text{obs}(z)$ represents a square region centered at~$z$ of length and width equal to 0.1. The objective is now to find the location $z$ of the square region~$\Omega_\text{obs}(z)$ inside the domain~$\Omega$ such that the influence matrix of~$h_\text{obs}$ matches as closely as possible the influence matrix of the QoIs at the validation scenario. Instead of solving directly the optimization problem~\eqref{eq:obj_function_sensor}, we plot the objective function in~\eqref{eq:obj_function_sensor} for various positions of the squared region inside~$\Omega$. We actually consider for the squared regions the grid cells shown in Figure~\ref{fig:domain}. Figure~\ref{fig:sensor_pollutant} presents the results for both QoIs under their respective optimal validation scenario. The best positions for the sensor are indicated in this figure by the squares in which the value of the objective function is the closest to zero.

For the first QoI, we clearly note that it is best to position the sensor within~$\Omega_1$, which is an obvious result since we ultimately seek to predict the mean concentration of  pollutant in~$\Omega_1$. The optimal region of observation seems to extend to the region where recirculation occurs, as shown in Figure~\ref{fig:velocity_field}. To some extent, it also spreads upstream above the first dock and downstream right above the second dock. Moreover, we observe that positioning the sensor in the upper-half of the domain would be useless for the prediction of the first QoI. This result is in fact quite intuitive since, under the optimal validation scenario shown in Figure~\ref{fig:pollutant_scenario}, very little pollutant flows into those regions while a certain amount of pollutant indeed reaches~$\Omega_1$. Finally, the influence matrices associated with the mean concentration in the regions located after the second dock share some similarities with the influence matrix~$M_{h_{qoi,1}}(x_\text{val})$, but not as much as the region between the two docks. We also remark that the fluctuations in the objective function in the plume located downstream could be explained by the accumulation of numerical errors resulting from the computation of the gradient (see~\ref{ann:gradient}).

For the second QoI, it is again optimal to position the sensor inside the region of interest, $\Omega_2$ in this case. The pollutant sensor can also be placed optimally right upstream and right downstream of~$\Omega_2$. Interestingly, it does not seem optimal to put the sensor right inside the plume of the pollutant. One should rather probe the boundary region of the plume next to the QoI since variations in the concentration are larger there and the measurements would thus be more sensitive to the location of the pollutant release.

\section{Conclusion}
\label{sec:conclusion}

We have addressed in this paper the issue of designing optimal validation experiments tailored toward the prediction of a QoI. To the best of our knowledge, this problem has been rarely considered in the literature. We thus provide a careful description of a mathematical model, the role of the various input parameters, and the objectives when performing predictions using the model. We have also carried out a consistent treatment of the aleatory uncertainty affecting the model parameters, the experimental and computed observations, and the QoI. Within this framework, we have introduced the influence matrix, computed using the Active Subspace method,  as a means to provide a quantitative description of the response surfaces associated with the various model functionals. We have then proposed a methodology to design an optimal validation scenario and identify optimal observations based on the comparison of influence matrices. The methodology essentially consists in the solution of two optimization problems: 1) the first optimal design problem~\eqref{eq:obj_function_control} allows the computation of the optimal validation scenario; 2) the second optimal problem~\eqref{eq:obj_function_sensor} finds the best measurements to be performed on the system of interest. 

The methodology was tested on two numerical examples. The results for a simple projectile problem demonstrated that the optimal validation scenario was able to recover the essential features of the influence matrix associated with the QoI at the prediction scenario. The influence matrices obtained with two different observable functionals were compared over a given time period and allowed us to identify an optimal observation functional and sensor parameter. The results for the pollutant transport problem highlighted the fact that the optimal validation experiment does indeed depend on the QoI we wish to predict. They confirmed as well the fact that, in the case of non-observable QoI, one can still identify observation functionals sharing similar influence matrices. This numerical example also shows that the proposed methodology can be viably used for applications of engineering interest. 

Several hypothesis and simplifications have been made throughout this work, limiting to some extent the scope of the predictive problem. Alleviating the hypothesis that the control parameters and sensor parameters are not uncertain could lead to more robust validation experiments, especially when these control parameters consist of boundary and/or initial conditions. Also, the use of the Active Subspace method to define the influence matrix necessitates some regularity on the observation and QoI functionals. For models lacking regularity, one could envision the use of the  \textit{variogram analysis of response surfaces}~\cite{razavi_new_2016} as a means to compute the influence matrix.

On another note, the study of the projectile problem hints that a sole dimensional analysis is insufficient to select an optimal validation experiment. However, transforming the parameter space with a dimensional analysis may yield a different representation of the influence matrix of a model functional. This new representation may provide additional information about the model. For example, we may be able to identify the important dimensionless quantities that guide the design of validation experiments. This particular question is the subject of ongoing research.

Moreover, in addition to employing the sensitivity analysis to design validation experiments, it could be used for the analysis of existing data sets. By describing the influence matrix of a model under a data set, one could for instance find specific data within this set that could be utilized for validation purposes. Adversarial validation scenarios aimed at testing a wide range of prediction settings may also be found. In the same vein, it could be insightful to apply the influence matrix to cross-validation approaches.

Finally, methods to quantify the model errors rely on some sort of comparison between the reality and the model observables. It could be interesting to analyze the impact of the choice of the experiment on the modeling error for the QoIs at the prediction scenario. The proposed methodology could perhaps better inform  the modeling error, which we will investigate in a future research.

\section*{Acknowledgements}
APR is grateful for the financial support of the \textit{Fonds de recherche du Québec - Nature et technologies}. SP and ML are grateful for the support from the Natural Sciences and Engineering Research Council of
Canada (NSERC) Discovery Grants [grant numbers RGPIN-2019-7154, RGPIN-2018-06592].


\bibliographystyle{abbrv}
\bibliography{bibliography_article_bibtex}

\begin{thebibliography}{10}

\bibitem{alnaes_fenics_2015}
M.~Alnæs, J.~Blechta, J.~Hake, A.~Johansson, B.~Kehlet, A.~Logg,
  C.~Richardson, J.~Ring, M.~Rognes, and G.~Wells.
\newblock The {FEniCS} project version 1.5.
\newblock {\em Archive of Numerical Software}, 3, 2015.

\bibitem{alnaes_unified_2014}
M.~S. Alnæs, A.~Logg, K.~B. Ølgaard, M.~E. Rognes, and G.~N. Wells.
\newblock Unified {Form} {Language}: {A} {Domain}-{Specific} {Language} for
  {Weak} {Formulations} of {Partial} {Differential} {Equations}.
\newblock {\em ACM Trans. Math. Softw.}, 40(2), 2014.

\bibitem{ao_design_2017}
D.~Ao, Z.~Hu, and S.~Mahadevan.
\newblock Design of validation experiments for life prediction models.
\newblock {\em Reliability Engineering \& System Safety}, 165:22--33, 2017.

\bibitem{arendt_quantification_2012}
P.~D. Arendt, D.~W. Apley, and W.~Chen.
\newblock Quantification of {Model} {Uncertainty}: {Calibration}, {Model}
  {Discrepancy}, and {Identifiability}.
\newblock {\em Journal of Mechanical Design}, 134(10), 2012.

\bibitem{atkinson_optimal_2015}
A.~C. Atkinson.
\newblock Optimal {Design}.
\newblock In {\em Wiley {StatsRef}: {Statistics} {Reference} {Online}}, pages
  1--17. John Wiley \& Sons, Ltd, 2015.

\bibitem{atkinson_optimum_2007}
A.~C. Atkinson, A.~N. Donev, and R.~D. Tobias.
\newblock {\em Optimum experimental designs, with {SAS}}.
\newblock Number~34 in Oxford statistical science series. Oxford University
  Press, 2007.

\bibitem{audet_algorithm_2022}
C.~Audet, S.~L. Digabel, V.~R. Montplaisir, and C.~Tribes.
\newblock Algorithm 1027: {NOMAD} {Version} 4: {Nonlinear} {Optimization} with
  the {MADS} {Algorithm}.
\newblock {\em ACM Transactions on Mathematical Software}, 48(3):22, 2022.

\bibitem{beck_fast_2018}
J.~Beck, B.~M. Dia, L.~F. Espath, Q.~Long, and R.~Tempone.
\newblock Fast {Bayesian} experimental design: {Laplace}-based importance
  sampling for the expected information gain.
\newblock {\em Computer Methods in Applied Mechanics and Engineering},
  334:523--553, 2018.

\bibitem{cole_parameter_2020}
D.~Cole.
\newblock {\em Parameter redundancy and identifiability}.
\newblock CRC Press, 2020.

\bibitem{constantine_active_2015}
P.~G. Constantine.
\newblock {\em Active subspaces: emerging ideas for dimension reduction in
  parameter studies}.
\newblock Number~2 in {SIAM} spotlights. Society for Industrial and Applied
  Mathematics, 2015.

\bibitem{constantine_active_2014}
P.~G. Constantine, E.~Dow, and Q.~Wang.
\newblock Active subspace methods in theory and practice: applications to
  kriging surfaces.
\newblock {\em SIAM Journal on Scientific Computing}, 36(4):A1500--A1524, 2014.
\newblock arXiv: 1304.2070.

\bibitem{da_veiga_basics_2021}
S.~Da~Veiga, F.~Gamboa, B.~Iooss, and C.~Prieur.
\newblock {\em Basics and {Trends} in {Sensitivity} {Analysis}}.
\newblock Computational {Science} \& {Engineering}. Society for Industrial and
  Applied Mathematics, 2021.

\bibitem{dasgupta_asymptotic_2008}
A.~DasGupta.
\newblock {\em Asymptotic {Theory} of {Statistics} and {Probability}}.
\newblock Springer {Texts} in {Statistics}. Springer New York, 2008.

\bibitem{farrell_bayesian_2015}
K.~Farrell, J.~T. Oden, and D.~Faghihi.
\newblock A {Bayesian} framework for adaptive selection, calibration, and
  validation of coarse-grained models of atomistic systems.
\newblock {\em Journal of Computational Physics}, 295:189--208, 2015.

\bibitem{ferson_model_2008}
S.~Ferson, W.~L. Oberkampf, and L.~Ginzburg.
\newblock Model validation and predictive capability for the thermal challenge
  problem.
\newblock {\em Validation Challenge Workshop}, 197(29):2408--2430, 2008.

\bibitem{goodman_ensemble_2010}
J.~Goodman and J.~Weare.
\newblock Ensemble samplers with affine invariance.
\newblock {\em Communications in Applied Mathematics and Computational
  Science}, 5(1):65--80, 2010.

\bibitem{hamilton_relation_2010}
J.~R. Hamilton and R.~G. Hills.
\newblock Relation of {Validation} {Experiments} to {Applications}.
\newblock {\em Numerical Heat Transfer, Part B: Fundamentals}, 57(5):307--332,
  2010.

\bibitem{hills_roll-up_2013}
R.~G. Hills.
\newblock Roll-up of validation results to a target application.
\newblock Technical Report SAND2013-7424, Sandia National Laboratories, United
  States, 2013.

\bibitem{hills_statistical_2003}
R.~G. Hills and I.~H. Leslie.
\newblock Statistical {Validation} of {Engineering} and {Scientific} {Models}:
  {Validation} {Experiments} to {Application}.
\newblock Technical Report SAND2003-0706, Sandia National Laboratories, United
  States, 2003.

\bibitem{hinze_optimization_2009}
M.~Hinze, R.~Pinnau, M.~Ulbrich, and S.~Ulbrich, editors.
\newblock {\em Optimization with {PDE} constraints}.
\newblock Number~23 in Mathematical modelling: theory and applications.
  Springer, 2009.

\bibitem{kennedy_bayesian_2001}
M.~C. Kennedy and A.~O'Hagan.
\newblock Bayesian calibration of computer models.
\newblock {\em Journal of the Royal Statistical Society: Series B (Statistical
  Methodology)}, 63(3):425--464, 2001.

\bibitem{klir_uncertainty_2006}
G.~J. Klir.
\newblock {\em Uncertainty and information: foundations of generalized
  information theory}.
\newblock Wiley-Interscience, 2006.

\bibitem{lee_review_2019}
G.~Lee, W.~Kim, H.~Oh, B.~D. Youn, and N.~H. Kim.
\newblock Review of statistical model calibration and validation—from the
  perspective of uncertainty structures.
\newblock {\em Structural and Multidisciplinary Optimization},
  60(4):1619--1644, 2019.

\bibitem{li_role_2016}
C.~Li and S.~Mahadevan.
\newblock Role of calibration, validation, and relevance in multi-level
  uncertainty integration.
\newblock {\em Reliability Engineering \& System Safety}, 148:32--43, 2016.

\bibitem{long_fast_2013}
Q.~Long, M.~Scavino, R.~Tempone, and S.~Wang.
\newblock Fast estimation of expected information gains for {Bayesian}
  experimental designs based on {Laplace} approximations.
\newblock {\em Computer Methods in Applied Mechanics and Engineering},
  259:24--39, 2013.

\bibitem{montoison_nomadjl_2020}
A.~Montoison, P.~Pascal, and L.~Salomon.
\newblock {NOMAD}.jl: {A} {Julia} interface for the constrained blackbox solver
  {NOMAD}, 2020.

\bibitem{mullins_separation_2016}
J.~Mullins, Y.~Ling, S.~Mahadevan, L.~Sun, and A.~Strachan.
\newblock Separation of aleatory and epistemic uncertainty in probabilistic
  model validation.
\newblock {\em Reliability Engineering \& System Safety}, 147:49--59, 2016.

\bibitem{mullins_survey_2016}
J.~G. Mullins, B.~B. Schroeder, R.~G. Hills, and L.~G. Crespo.
\newblock A {Survey} of {Methods} for {Integration} of {Uncertainty} and
  {Model} {Form} {Error} in {Prediction}.
\newblock In {\em Report {Number}: {SAND2016}-{4585C}}, 2016.
\newblock Sandia National Laboratories, United States.

\bibitem{oberkampf_verification_2008}
W.~L. Oberkampf and T.~G. Trucano.
\newblock Verification and validation benchmarks.
\newblock {\em Benchmarking of CFD Codes for Application to Nuclear Reactor
  Safety}, 238(3):716--743, 2008.

\bibitem{oliver_validating_2015}
T.~A. Oliver, G.~Terejanu, C.~S. Simmons, and R.~D. Moser.
\newblock Validating predictions of unobserved quantities.
\newblock {\em Computer Methods in Applied Mechanics and Engineering},
  283:1310--1335, 2015.

\bibitem{razavi_new_2016}
S.~Razavi and H.~V. Gupta.
\newblock A new framework for comprehensive, robust, and efficient global
  sensitivity analysis: 1. {Theory}.
\newblock {\em Water Resources Research}, 52(1):423--439, 2016.

\bibitem{razavi_future_2021}
S.~Razavi, A.~Jakeman, A.~Saltelli, C.~Prieur, B.~Iooss, E.~Borgonovo,
  E.~Plischke, S.~Lo~Piano, T.~Iwanaga, W.~Becker, S.~Tarantola, J.~H.
  Guillaume, J.~Jakeman, H.~Gupta, N.~Melillo, G.~Rabitti, V.~Chabridon,
  Q.~Duan, X.~Sun, S.~Smith, R.~Sheikholeslami, N.~Hosseini, M.~Asadzadeh,
  A.~Puy, S.~Kucherenko, and H.~R. Maier.
\newblock The {Future} of {Sensitivity} {Analysis}: {An} essential discipline
  for systems modeling and policy support.
\newblock {\em Environmental Modelling \& Software}, 137:104954, 2021.

\bibitem{rebba_computational_2008}
R.~Rebba and S.~Mahadevan.
\newblock Computational methods for model reliability assessment.
\newblock {\em Reliability Engineering \& System Safety}, 93(8):1197--1207,
  2008.

\bibitem{riedmaier_unified_2021}
S.~Riedmaier, B.~Danquah, B.~Schick, and F.~Diermeyer.
\newblock Unified {Framework} and {Survey} for {Model} {Verification},
  {Validation} and {Uncertainty} {Quantification}.
\newblock {\em Archives of Computational Methods in Engineering},
  28(4):2655--2688, 2021.

\bibitem{roy_comprehensive_2011}
C.~J. Roy and W.~L. Oberkampf.
\newblock A comprehensive framework for verification, validation, and
  uncertainty quantification in scientific computing.
\newblock {\em Computer Methods in Applied Mechanics and Engineering},
  200(25):2131--2144, 2011.

\bibitem{ryan_review_2016}
E.~G. Ryan, C.~C. Drovandi, J.~M. McGree, and A.~N. Pettitt.
\newblock A {Review} of {Modern} {Computational} {Algorithms} for {Bayesian}
  {Optimal} {Design}.
\newblock {\em International Statistical Review}, 84(1):128--154, 2016.

\bibitem{saltelli_global_2007}
A.~Saltelli, M.~Ratto, T.~Andres, F.~Campolongo, J.~Cariboni, D.~Gatelli,
  M.~Saisana, and S.~Tarantola.
\newblock {\em Global {Sensitivity} {Analysis}. {The} {Primer}}.
\newblock John Wiley \& Sons, Ltd, 2007.

\bibitem{sivia_data_2006}
D.~S. Sivia and J.~Skilling.
\newblock {\em Data analysis: a {Bayesian} tutorial}.
\newblock Oxford science publications. Oxford University Press, 2nd ed edition,
  2006.

\bibitem{sobol_global_2001}
I.~Sobol.
\newblock Global sensitivity indices for nonlinear mathematical models and
  their {Monte} {Carlo} estimates.
\newblock {\em Mathematics and Computers in Simulation}, 55(1):271--280, 2001.

\bibitem{sullivan_introduction_2015}
T.~Sullivan.
\newblock {\em Introduction to {Uncertainty} {Quantification}}, volume~63 of
  {\em Texts in {Applied} {Mathematics}}.
\newblock Springer International Publishing, 2015.

\bibitem{sunseri_hyper-differential_2020}
I.~Sunseri, J.~Hart, B.~van Bloemen~Waanders, and A.~Alexanderian.
\newblock Hyper-differential sensitivity analysis for inverse problems
  constrained by partial differential equations.
\newblock {\em Inverse Problems}, 36(12):125001, 2020.

\bibitem{tan_toward_2022}
J.~Tan, B.~Liang, P.~K. Singh, K.~A. Farrell-Maupin, and D.~Faghihi.
\newblock Toward selecting optimal predictive multiscale models.
\newblock {\em Computer Methods in Applied Mechanics and Engineering}, 402,
  2022.

\bibitem{white_fluid_2009}
F.~M. White.
\newblock {\em Fluid mechanics}.
\newblock Mcgraw-{Hill} series in mechanical engineering. McGraw-Hill, 7th ed
  edition, 2009.

\bibitem{yang_general_2015}
H.~Yang, M.~Fan, A.~Liu, and L.~Dong.
\newblock General formulas for drag coefficient and settling velocity of sphere
  based on theoretical law.
\newblock {\em International Journal of Mining Science and Technology},
  25(2):219--223, 2015.

\end{thebibliography}

\appendix


\section{Velocity Field for the Pollutant Transport Example} \label{ann:velocity}

\begin{figure}[t]
    \centering
    \setlength\figureheight{0.25\linewidth}
    \setlength\figurewidth{0.05\linewidth}
    \setlength{\tabcolsep}{5pt}
    \begin{tabular}{rl}
        \tikzset{external/export next=false}
    \begin{tikzpicture}[baseline,scale=\figureheight/2cm]
        \node[inner sep=0pt,anchor=south west,transform shape] (origin) at (0,0)
        {\includegraphics[height=2cm,width=5cm]{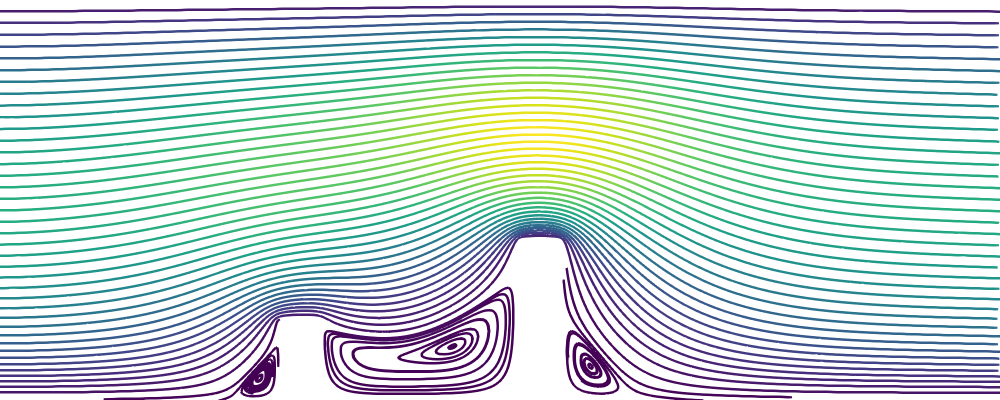}};

        \node [coordinate] (origin) at (0,0)  {};
        \node [coordinate] (v2) at (1.4,0)  {};
        \node [coordinate] (v4) at (1.6,0.4)  {};
        
        \node [coordinate] (v6) at (2.6,0)  {};
        \node [coordinate] (v8) at (2.8,0.8)  {};
        \node [coordinate] (v11) at (5,2)  {};
        
        \draw [draw=black] (origin) -- (v2)  -- 
        (1.4,0.4) -- 
        (v4) -- 
        (1.6,0) -- 
        (v6) -- 
        (2.6,0.8) -- 
        (v8) -- 
        (2.8,0) -- 
        (5,0) -- 
        (v11) -- 
        (0,2) -- 
        cycle;
            
    \end{tikzpicture}
        &
        \tikzset{external/export next=false}
        \colorbarcustom{0}{1.6}
    \end{tabular}
    \caption{Streamline of the velocity field $\vel$}
    \label{fig:velocity_field}
\end{figure}

The velocity field~$v$ for the pollutant transport problem is obtained by solving the Stokes' equation with the following boundary conditions
\begin{subequations} 
\label{eq:velocity_model_contaminant}
\begin{align}
    - \nu \Delta \vel + \nabla \beta = 0, &\qquad \text{in}\
    \Omega,\\
    \nabla \cdot \vel = 0, &\qquad \text{in}\ \Omega,\\
    \vel = 0, & \qquad \text{on}\ \Gamma_\text{banks}\cup\Gamma_\text{dock},\\
    \vel = \vel_D, &\qquad \text{on}\ \Gamma_\text{west},\\
    n \cdot (\nu \nabla \vel - \beta I) = 0,  &\qquad \text{on}\ \Gamma_\text{east},
\end{align}
\end{subequations}
where~$\nu\approx 0.001$ is approximately the dynamic viscosity of water and~$\beta$ is the pressure. The Dirichlet conditions $\vel_D$ consist in the fully developed quadratic flow 
\begin{equation}
    \vel_D(z) = \vel_\text{max} (2-z_2)z_2,
\end{equation}
where the maximal velocity $\vel_\text{max}$ is taken here to  be equal to unity. We solve the weak formulation of the problem with the open-source Finite Element code FEniCSx~\cite{alnaes_fenics_2015}. The streamlines of the flow are displayed in Figure~\ref{fig:velocity_field}.

We observe the presence of three recirculation zones, one before the first dock, a large one between the two docks, and another small one after the second dock. The choice of the Stokes model to represent the velocity field is only a coarse approximation for this application. However, the sole purpose of this model is to obtain a velocity field for the pollutant transport model. For example, if the velocity field were given by the solution to the Navier-Stokes equations, then the pollutant transport model~\eqref{eq:model_contaminant} would be different from the one under study. However, we would use the same methodology to obtain the optimal validation scenarios.

\section{Gradient for the Contaminant Transport Example} \label{ann:gradient}

The computation of the gradient (with respect to the parameters) of the QoI and the observables for the pollutant transport model described in Section~\ref{ssec:contaminant_problem} is somewhat involved. In order to do so, we compute the adjoint solution of the diffusion-advection equation~\eqref{eq:model_contaminant} and use it to obtain the derivatives~\cite{hinze_optimization_2009}. As previously mentioned, the state variables~$(\vel,\beta)$ for the velocity and pressure of the water can be solved independently from the advection-diffusion equation and do not vary with the parameters. 

We explicitly introduce a lift function~$\bar{\conc}_D$ to account for the Dirichlet boundary condition $\conc=\conc_D$, as we wish to compute the derivative of the QoI and observables with respect to the parameters that define~$\conc_D$, so~$\conc = \bar{\conc} + \bar{\conc}_D$. The weak form of the advection-diffusion reads: 
\begin{equation}
\begin{aligned}
& \text{Find~$\bar{\conc} \in V=\{\psi\in H^1(\Omega);\  \psi|_{\Gamma_\text{west}}=0\}$ such that}\\
& \underbrace{\int_\Omega \exp(k) \nabla \bar{\conc} \cdot \nabla \psi +  \nabla \cdot (\vel  \bar{\conc}) \psi\, \diff z}_{= a(\bar{\conc},v)} + \underbrace{\int_\Omega \exp(k) \nabla \bar{\conc}_D \cdot \nabla \psi + \nabla \cdot (\vel \bar{\conc}_D) \psi\, \diff z}_{=a(\bar{\conc}_D,\psi):=l(\psi)}=0, \quad \forall\ \psi \in V.
\end{aligned}
\end{equation}
The equilibrium relation for this problem reads:
\begin{align}
    r(x,\theta,\conc) = a(\conc,\cdot) =  a(\bar{\conc},\cdot) + a(\bar{\conc}_D,\cdot) \in V^*,
\end{align}
where~$V^*$ is the dual space of~$V$. From~\cite[Section 1.6]{hinze_optimization_2009}, the computation of the derivative is performed in two steps. We first solve the adjoint problem for $\lambda$ such that
\begin{align}
\label{eq:adjoint_problem}
    r_\conc(p,\conc(p))^* \lambda = - h_\conc(p,\conc(p)),
\end{align}
and then compute the gradient according to
\begin{align}
\label{eq:der_model_contaminant}
    \nabla_p h(p,\theta) = r_p\big(p,\conc(p) \big)^* \lambda + h_p\big(p,\conc(p) \big).
\end{align}
The subscripts~$\conc$ and~$p$ stands for the variables with respect to which we perform differentiation and~$^*$ denotes the adjoint. For the problem at hand, the partial derivative of the equilibrium relation in the direction~$\hat{\conc}$ is
\begin{align}
    r_\conc(p,\conc(p))\hat{\conc} = a(\hat{\conc},\cdot),
\end{align}
while its adjoint is
\begin{equation}
\begin{aligned}
    \dual{r_\conc(p,\conc(p))^* \psi,\conc}_{V^*,V} & = \dual{r_\conc(p,\conc(p))\conc,\psi}_{V^*,V} \\  
    & = a(\conc,\psi) \\
    & = \int_\Omega \exp(k) \nabla \conc \cdot \nabla \psi + \nabla \cdot (\vel \conc) \psi\, \diff z.
\end{aligned}
\end{equation}
Hence, the adjoint problem reads: \begin{equation}
\label{eq:adjoint_solution}
\begin{aligned}
& \text{Find~$\lambda \in V$ such that} \\
& \qquad  \int_\Omega \exp(k) \nabla \psi \cdot \nabla \lambda + \nabla \cdot (\vel \psi) \lambda\, \diff z = - \frac{1}{\abs{\omega}}\int_{\omega} \psi \, \diff z \, , \qquad \forall \, \psi \in V, 
\end{aligned}
\end{equation}

where~$\omega$ can be either~$\Omega_1$,~$\Omega_2$, or~$\Omega_\text{obs}$. We now need to compute the partial derivative of the equilibrium relation~$r(p,\conc(p))$ and of the QoI/observables~$h(p,\conc(p),z)$ with respect to the parameters~$p$. As a reminder, the control parameters for this problem are the location, width, and height for the Dirichlet condition, that is~$x=(z_0, L, c)$), respectively, and the model parameter is the diffusivity term~$\theta=k$. It is at this point that the lift function will be employed. By applying the chain rule, we obtain the partial derivative with respect to the control parameters~$x$ in the direction~$\hat{x}$
\begin{align}
    r_x(p,\conc(p))\hat{x} = r_{\bar{\conc}_D}(p,\conc(p)) \bar{\conc}_{D,x}(x,z)\hat{x} = a(\bar{\conc}_{D,x}(x,z) \cdot \hat{x},\cdot),
\end{align}
where~$\bar{\conc}_{D,x}$ denotes the gradient of~$\bar{\conc}_D$ with respect to~$x$. Its adjoint is computed as
\begin{equation}
\begin{split}       \dual{r_x(p,\conc(p))^*\psi,\hat{x}}_{\mathcal{X}^*,\mathcal{X}} &= \dual{r_x(p,\conc(p))\hat{x},\psi}_{V^*,V} = a(\bar{\conc}_{D,x}(x,z) \cdot \hat{x},\psi)\\
&= \int_\Omega \exp(k) \nabla (\bar{\conc}_{D,x}(x,z)\cdot \hat{x}) \cdot \nabla \psi + \nabla \cdot \big(\vel (\bar{\conc}_{D,x}(x,z)\cdot \hat{x})\big) \psi\, \diff z\\
&= \int_\Omega \Big(\exp(k) \nabla \bar{\conc}_{D,x}(x,z) \cdot \nabla \psi + \nabla \cdot(\vel \bar{\conc}_{D,x}(x,z)) \psi\Big)\cdot \hat{x}\, \diff z.
\end{split}
\end{equation}
So the gradient of the QoI/observable~$h$ with respect to the control parameters is given by
\begin{align}
\label{eq:derivative_control_parameter}
    \nabla_x h(p,\conc(p)) = \int_\Omega \exp(k) \nabla \bar{\conc}_{D,x}(x,z) \cdot \nabla \lambda + \nabla \cdot (\vel \bar{\conc}_{D,x}(x,z)) \lambda \, \diff z + 
    \frac{1}{\abs{\omega}}\int_{\omega} \bar{\conc}_{D,x}(x,z) \, \diff z.
\end{align}
The derivative with respect to the model parameter~$\theta=k$ can be similarly computed as
\begin{align}
\label{eq:derivative_model_parameter}
    \nabla_\theta h(p,\conc(p)) = \int_\Omega \exp(k) \nabla \conc \cdot \nabla \lambda \, \diff z. 
\end{align}
The adjoint solution~\eqref{eq:adjoint_solution}, the gradient~\eqref{eq:derivative_control_parameter}, and the derivative~\eqref{eq:derivative_model_parameter} are evaluated within the Finite Element code FEniCSx~\cite{alnaes_fenics_2015} using the module UFL~\cite{alnaes_unified_2014}.

\section{Solution of the Optimal Design Problems} 
\label{ann:optimization}

The optimization problems~\eqref{eq:obj_function_control} are solved with the solver NOMAD~\cite{audet_algorithm_2022} using the implementation of the NOMAD code~\cite{montoison_nomadjl_2020} in the programming language Julia. The main reason is that NOMAD is a derivative-free solver, meaning that only the value of the objective function is needed. The computation of the derivative of the objective function in problems~\eqref{eq:obj_function_control} and~\eqref{eq:obj_function_sensor} requires computing the Hessian of the observation and QoI functionals~$h_\text{obs}$ and~$h_\text{qoi}$ with respect to the control and model parameters~$x$ and~$\theta$. Their calculation is often too  computationally intensive, so we prefer here to rely on a derivative-free optimization solver. NOMAD allows one to find local minima, but additional resources can be allocated to perform a global search. This option is employed in this work for reasons that are outlined below. 

As mentioned in Section~\ref{ssec:design_validation_scenario}, the objective functions in Problems~\eqref{eq:obj_function_control} and~\eqref{eq:obj_function_sensor} may not possess a unique global minimum. Moreover, the objective functions may possess several local minima. Hence, it is beneficial to perform a global search in the controlled environment~$\mathcal{X}_\text{lab}$ and constrained set~$\mathcal{Z}_\text{lab}$ to escape the basin of attraction of a local minimum and hopefully find a global minimum.

To gain insight into the behavior of the objective function of Problem~\eqref{eq:obj_function_control} and to assess the performance of NOMAD, we perform two verification tests. These two verification tests consist in solving~\eqref{eq:obj_function_control} for the two models described in Sections~\ref{ssec:example1} and~\ref{ssec:contaminant_problem}, but for a controlled environment~$\mathcal{X}_\text{lab}$ now equal to the control space~$\mathcal{X}$. A global minimum is guaranteed to exist in that case, which precisely corresponds to the prediction scenario. For the projectile model, the control parameters space is~$\mathcal{X} = [0.005,5] \times [0.005,0.1] \times[0,2] \times [10,120]$. As described earlier, the prediction scenario is~$x_\text{pred} = (0.05,0.01,1,100)$. We initialize the optimization algorithm with ten different initial points distributed according to the Latin Hypercube algorithm. Table~\ref{tab:verification_free_falling} reports the local minimum attained for each initial point. 
The values of the objective function normalized with respect to~$\norm*{M_{h_\text{qoi}}(x_\text{pred})}$ indicate that the quality of all local minima is similar. However, when we look at the error, measured as the~$l_2$ distance between the computed optimal point and the prediction scenario~$x_\text{pred}$, we observe that no initial point leads to the prediction scenario. This result can be explained by analyzing more closely the optimal points. On one hand, all have succeeded in recovering the value of the mass~$m$, diameter~$\ell$, and initial velocity~$v_0$ of the prediction scenario~$x_\text{pred}$. On the other hand, values obtained for the initial position~$u_0$ are somewhat different. This result arises from the fact that the QoI~\eqref{eq:qoi_projectile} is affine with respect to the initial position~$u_0$. Hence, the influence matrix is the same whatever the value of the initial position~$u_0$ (for fixed value of the other control parameters), and therefore reflected by the values of the objective function. We may conclude that a unique global minimum does not exist for this problem. However, there exists a one-dimensional linear manifold on which the global minimum can be attained.

\begin{table}[t]
\centering
\caption{Results of the ten verification experiments for the projectile problem}
\vspace{2pt}
\begin{tabular}{@{}ccccccccccc@{}}
\bottomrule
 & \multicolumn{4}{c}{\begin{tabular}[c]{@{}c@{}} \\ Initial Point\end{tabular}} & \multicolumn{4}{c}{\begin{tabular}[c]{@{}c@{}} \\ Optimal Point\end{tabular}} & \multirow{3}{*}{Error} & \multirow{3}{*}{\begin{tabular}[c]{@{}c@{}} Normalized\\ Objective\\ Function\end{tabular}} \\ \cmidrule(lr){2-5} \cmidrule(lr){6-9}
                  & $m$    & $\ell$  & $u_0$  & $v_0$ & $m$  & $\ell$   & $u_0$   & $v_0$ &                        &                                                                                 \\ \midrule
1                 & 2.535  & 0.02359 & 0.8801 & 76.87 & 0.05 & 0.009999 & 1.126   & 100   & 0.1263                 & 1.6e-05                                                                         \\
2                 & 1.438  & 0.02451 & 0.4384 & 56.95 & 0.05 & 0.01     & 0.5074  & 100   & 0.4926                 & 1e-06                                                                           \\
3                 & 0.7639 & 0.05092 & 0.1344 & 43.16 & 0.05 & 0.01     & 1.752   & 100   & 0.7515                 & 1e-06                                                                           \\
4                 & 1.721  & 0.05431 & 1.291  & 112.5 & 0.05 & 0.01     & 0.8704  & 100   & 0.1296                 & 1e-06                                                                           \\
5                 & 4.257  & 0.04152 & 0.2073 & 97.81 & 0.05 & 0.009999 & 0.41    & 100   & 0.59                   & 4e-06                                                                           \\
6                 & 3.573  & 0.08897 & 0.6992 & 106.9 & 0.05 & 0.01     & 1.977   & 100   & 0.9768                 & 1e-06                                                                           \\
7                 & 4.754  & 0.0655  & 1.663  & 33.4  & 0.05 & 0.01     & 1.703   & 100   & 0.7028                 & 1e-06                                                                           \\
8                 & 2.149  & 0.01172 & 1.061  & 27.74 & 0.05 & 0.01     & 0.06525 & 100   & 0.9348                 & 1e-06                                                                           \\
9                 & 0.3243 & 0.09373 & 1.96   & 18.72 & 0.05 & 0.01     & 1.23    & 100   & 0.23                   & 1e-06                                                                           \\
10                & 3.14   & 0.07903 & 1.491  & 75.87 & 0.05 & 0.01     & 1.606   & 100   & 0.6062                 & 1.6e-05                                                                         \\ \bottomrule
\end{tabular}
\label{tab:verification_free_falling}
\end{table}

We repeat the same verification process for the pollutant transport model. The prediction scenario is~$x_\text{pred}=(0.75,0.6,2)$ and the control parameters space is~$\mathcal{X} = [0.1,1.9]\times[0.005,1]\times[0.1,10]$. Table~\ref{tab:verification_pollutant_transport} provides the solutions obtained using NOMAD for each initial control parameter. The verification experiments number 3, 9, and 10 seem to indicate the existence of a local minima away for the prediction scenario. However, the values of their normalized objective function indicate that the influence matrices of the QoI at the optimal points and at the prediction scenario~$x_\text{pred}$ are similar. The other verification experiments possess lower value of normalized objective function, even if some optimal points are not very close to the prediction point.

\begin{table}[t]
\centering
\caption{Results of the ten verification experiments for the pollutant transport problem}
\vspace{2pt}
\begin{tabular}{@{}ccccccccc@{}}
\bottomrule
 & \multicolumn{3}{c}{\begin{tabular}[c]{@{}c@{}} \\ Initial Point\end{tabular}} & \multicolumn{3}{c}{\begin{tabular}[c]{@{}c@{}} \\ Optimal Point\end{tabular}} & \multirow{3}{*}{Error} & \multirow{3}{*}{\begin{tabular}[c]{@{}c@{}} Normalized \\ Objective\\ Function\end{tabular}} \\ \cmidrule(lr){2-4} \cmidrule(lr){5-7}
                  & $z_0$    & $L$  & $c$ & $z_0$    & $L$  & $c$ &                        &                                                                                 \\ \midrule
1  & 0.9751   & 0.9485  & 5.141      & 0.7445   & 0.5915  & 1.994      & 0.01178 & 0.000178 \\
2  & 1.057    & 0.8015  & 1.978      & 0.7001   & 0.5231  & 1.95       & 0.1047  & 0.00052  \\
3  & 1.275    & 0.7156  & 6.788      & 0.9793   & 0.9626  & 2.327      & 0.5392  & 0.01186  \\
4  & 0.6539   & 0.6264  & 0.9501     & 0.8002   & 0.6802  & 2.059      & 0.1115  & 0.002432 \\
5  & 0.541    & 0.5098  & 7.535      & 0.7852   & 0.6554  & 2.041      & 0.07719 & 0.00158  \\
6  & 1.496    & 0.4369  & 2.966      & 0.6955   & 0.5164  & 1.946      & 0.1136  & 0.00011  \\
7  & 1.651    & 0.3441  & 9.502      & 0.8258   & 0.7204  & 2.092      & 0.1696  & 0.003887 \\
8  & 0.3233   & 0.2538  & 8.832      & 0.7561   & 0.6098  & 2.007      & 0.0133  & 0.000274 \\
9  & 1.832    & 0.1114  & 3.718      & 0.9329   & 0.8938  & 2.253      & 0.4287  & 0.009839 \\
10 & 0.1281   & 0.04033 & 4.255      & 0.9048   & 0.8467  & 2.207      & 0.3571  & 0.008336                                                                         \\ \bottomrule
\end{tabular}
\label{tab:verification_pollutant_transport}
\end{table}

These verification experiments demonstrate the robustness of the solver NOMAD for solving the optimal design problem~\eqref{eq:obj_function_control}. We are thus confident that the optimal validation scenarios obtained earlier for the projectile and pollutant transport problems are reasonable local minima of the objective function~\eqref{eq:obj_function_control} under their respective constrained sets~$\mathcal{X}_\text{lab}$.

\end{document}